\documentclass[12pt]{article}

\usepackage[dvips]{graphicx}
\input rus.def
\font\scaps=cmcsc10    

\newcommand \be{\begin{equation}}
\newcommand \ee{\end{equation}}
\newcommand \ba{\begin{eqnarray}}
\newcommand \ea{\end{eqnarray}}
\begin{document}

\def\today{\ifcase\month\or
 January\or February\or March\or April\or May\or June\or
 July\or August\or September\or October\or November\or December\fi
 \space\number\day, \number\year}
%

\hfil PostScript file created: \today{}; \ time \the\time \ minutes
\vskip .15in

%



\centerline {EARTHQUAKE SIZE DISTRIBUTION: POWER-LAW WITH
EXPONENT $\beta \equiv {1 \over 2} \ ?$
}

\vskip .15in
\begin{center}
{Yan Y. Kagan }
\end{center}
\centerline {Department of Earth and Space Sciences,
University of California,}
\centerline {Los Angeles, California 90095-1567, USA;}
\centerline {Emails: {\tt ykagan@ucla.edu,
kagan@moho.ess.ucla.edu }}
\vskip 0.02 truein

\vspace{0.15in}

\noindent
{\bf Abstract.}
We propose that the widely observed and universal
Gutenberg-Richter relation is a mathematical consequence of
the critical branching nature of earthquake process in a
brittle fracture environment.
These arguments, though preliminary, are confirmed by
recent investigations of the seismic moment distribution in
global earthquake catalogs and by the results on the
distribution in crystals of dislocation avalanche sizes.
We consider possible systematic and random errors in
determining earthquake size, especially its seismic moment.
These effects increase the estimate of the parameter $\beta$
of the power-law distribution of earthquake sizes.
In particular, we find that estimated $\beta$-values may be
inflated by 1-3\% because relative moment uncertainties
decrease with increasing earthquake size.
Moreover, earthquake clustering greatly influences the
$\beta$-parameter.
If clusters (aftershock sequences) are taken as the
entity to be studied, then the exponent value for their size
distribution would decrease by 5-10\%.
The complexity of any earthquake source also inflates the
estimated $\beta$-value by at least 3-7\%.
The centroid depth distribution also should influence the
$\beta$-value, an approximate calculation suggests that the
exponent value may be increased by 2-6\%.
Taking all these effects into account, we propose that the
recently obtained $\beta$-value of 0.63 could be reduced to
about 0.52--0.56: near the universal constant value (1/2)
predicted by theoretical arguments.
We also consider possible consequences of the universal
$\beta$-value and its relevance for theoretical and practical
understanding of earthquake occurrence in various tectonic and
Earth structure environments.
Using comparative crystal deformation results may help us
understand the generation of seismic tremors and slow
earthquakes and illuminate the transition from brittle
fracture to plastic flow.

\vskip .15in
\noindent
{\bf Short running title}:
{\sc
Earthquake size distribution
}


\vskip .25in

\section{Introduction}
\label{intro}

Earthquake size distribution is usually described by the
Gutenberg-Richter (G-R) magnitude-frequency relation (see
Section~\ref{ana1}).
The G-R distribution can be transformed into the Pareto
(power-law) distribution for a scalar seismic moment $M$ with
the exponent $\beta = b/1.5$, where $b$ is the parameter of
the G-R law.

Theoretical analysis of earthquake occurrence (Vere-Jones,
1976, 1977) suggests that, given its branching nature, the
exponent $\beta$ of earthquake size distribution should be
identical to 1/2.
Properties of the critical branching process explain this
result: the total number of events (individuals) in such a
process asymptotically is distributed according to a power-law
with exponent 0.5 (Otter, 1949; Harris, 1963, Ch.\ I.13).
Such distributions, obtained by simulations, are shown as
upper curves in Figs.~\ref{fig7} and \ref{fig8} below.

The same values of power-law exponents are derived for
percolation and self-organized critcality (SOC) processes in a
high-dimensional space (see discussion by Kagan, 1991a,
p.~132).
Similar values of exponents are obtained by theoretical
arguments and simulations for dislocation avalanches in
crystalline materials (Zaiser, 2006, and references therein).

However, almost all the $\beta$ or $b$ measurements in
earthquake catalogs result in estimates larger than ${1 \over
2}$ (or 0.75 for the $b$-value).
These estimates exhibit large variations in different regions,
tectonic zones, etc.
A search of the ISI database (Thomson Reuters Scientific)
indicates that over the last five years, more than three
papers on the $b$-value or the G-R relation were published
monthly.
The similar rate of publications of the papers discussing the
magnitude-frequency relation can be observed in the previous
20-30 years.
In almost all articles the variation of the $b$-value is
attributed to different tectonics, rock stress, etc.

The following reasons for variability in the measured $b$-
and $\beta$-values can be proposed:

$\bullet$
1. Inappropriate usage of magnitude scales other than moment
magnitude: only the moment magnitude should be studied.
Regular earthquake magnitudes have significant systematic and
random errors (Kagan, 1999, 2003), making them inappropriate
for rigorous statistical, quantitative investigation.

$\bullet$
2. The maximum or corner moment $M_c$ (see Section~\ref{ana1})
needs careful consideration (Kagan, 1991a; 2002a; Bird and
Kagan, 2004).
If $M_c$ close to the magnitude/moment threshold -- that is,
the smallest magnitude/moment above which the catalog can be
considered to be complete -- and the earthquake size
distribution is approximated by a plain G-R distribution, the
magnitude-frequency curve would shift downwards, and the $b$-
or $\beta$-estimates would be strongly biased upwards.
In such a case to avoid the bias it is necessary to apply a
two-parameter relation, which includes the maximum or corner
magnitude.

$\bullet$
3. Mixing populations of earthquakes from tectonic settings
that have different corner magnitudes: when earthquake
populations with a varying corner moment $M_c$ are placed in
the study samples, a false increase in the $\beta$-values
results.
Mixing populations with different $M_c$ may even yield a
seemingly linear curve in a log-log plot, in such a case a
two-parameter approximation of the magnitude-frequency
relation would fail to avoid bias.
For example, due to a significant corner moment variability,
Kagan (2002a, Table~5, Section~5.2.3) determined the
$\beta$-values on the order 0.8-1.1 for mid-ocean earthquakes.
Similarly and apparently for the same reason, excessively large
$b$-values for oceanic earthquakes were obtained by Okal
and Romanowicz (1994) and Schorlemmer {\it et al.}\ (2005).

$\bullet$
4. Relative seismic moment errors increase with decreases in
earthquake size, resulting in a spurious $\beta$ increase.

$\bullet$
5. The object of study should be earthquake sequences, not
individual earthquakes: theoretical estimates, discussed
above, are relevant for earthquake {\sl sequences}, not
individual events.
Hence $\beta$-values need to be corrected for this effect.

$\bullet$
6. An earthquake is a tensor; its size, as given in moment
tensor solutions, is a tensor sum of earthquake subevents.
If these subevents have different focal mechanisms, their
tensor sum would be smaller than that for scalar moments of
subevents.
Thus, even if the number of elementary earthquake events were
distributed according to the power-law with the exponent
$\beta \equiv {1 \over 2}$, the distribution of earthquake
size, as represented by the tensor sum, may have a larger
exponent value.
This follows from stochastic complexity in the source.
If, for example, a source consists of positive and negative
random $n$ subsources, its size would be proportional to
$\sqrt n$: the size would be similar to that at the end
of a Brownian random walk.

$\bullet$
7. The corner moment $M_c$ is likely to change significantly
with the depth for shallow earthquakes (Kagan, 1999,
Section~1; Kagan, 2002a).
For shallow seismicity, earthquake populations with various
depths are usually added up, thus $\beta$ determinations are
biased.
Earthquake catalogs with only the hypocenter information
cannot be used to investigate this effect because hypocenters
are often located at a lower or upper depth boundary of a
fault rupture zone.
On the other hand, the degree of accuracy of centroid depths
in moment tensor catalogs is presently too low to study
rigorously the $M_c$ depth variation.

On the basis of statistical analysis of several earthquake
catalogs and some theoretical observations, Kagan (1991a,
pp.~129, 132-3) conjectured that the $\beta$-value is a
universal constant (1/2).
Three additional arguments can now be added to strengthen this
hypothesis:
(a) more careful measurements of the $\beta$ parameter in
modern earthquake catalogs (mainly the CMT catalog) suggest
that the $\beta$ is universal;
(b) recent advances in space geodesy and quantitative plate
tectonics allowed for detailed calculation of tectonic
deformation rate.
By comparing tectonic and seismic moment rates, we can
calculate the upper bound, $M_c$, for earthquake moment
distribution; and
(c) investigation of dislocation avalanches in crystals
indicated that their size distribution is a power-law whose
exponent has a universal value.

Several previous investigations (Kagan, 1999; Bird and Kagan,
2004; Boettcher and Jordan, 2004) suggested that the $\beta$
has a universal value on the order of 0.63--0.67.
Moreover, statistical analysis of earthquake size distribution
and comparison of seismic and tectonic deformation rates
allowed us to evaluate the corner moment ($M_c$) value for
several tectonic regions.
We conclude that apparent change in the $\beta$-values is due
mainly to $M_c$ variability.

However, the $\beta$ universality model can be challenged.
One can argue that more careful measurements may reveal
statistically significant variations in the $\beta$ exponent.
The aim here is to show that the $\beta$-value is a
universal constant.
We consider various systematic and random effects whose
influence would confirm this conjecture.

Recent experimental and theoretical investigations have
demonstrated that crystal plasticity is characterized by large
intrinsic spatio-temporal fluctuations with scale-invariant
characteristics.
In other words, deformation proceeds through intermittent
bursts with power-law size distributions (Zaiser, 2006;
Dahmen {\it et al.}, 2009).
In particular, Csikor {\it et al.}\ (2007), Dimiduk {\it et
al.}\ (2006), Weiss and Marsan (2003) studied
dislocation avalanches (micro- and nano-earthquakes) in ice
and other crystals.
They consistently obtain the power-law size distribution
with the probability density exponent close to 1.5.
This would correspond to our exponent $1+\beta$.
The power-size distribution in a single ice crystal extends
over six decades of magnitude (Miguel {\it et al.}, 2001).

Comparing these dislocation avalanche measurements with
seismological observations leads to some problems.
The most accurate measurements of earthquake size are for
the seismic moment.
Experimental laboratory observations provide the energy of
acoustic emission bursts or strain step measurements.
Zaiser (2006, pp.~212, 223) argues that both these
measurements are approximately equivalent and yield similar
values for the distribution density power-law exponent
(1.5--1.6).

Earthquake energy has the same distribution as the seismic
moment.
Kanamori (1977) cites the following relation between energy
released by earthquakes, $E$, and their magnitude, $m$:
$\log_{10} E \, = \, 1.5 \, m \, + \, 11.8$.
Because the moment and magnitudes have a similar dependence
(Kagan, 2002a), the exponents for energy and moment power-law
distributions should be identical.

Another problem of comparison is that obtained statistical
distributions of dislocation avalanches are not processed by
appropriate statistical techniques.
Thus, the obtained values of the exponents may be biased.
For example, Zaiser {\it et al.}\ (2008) use the least-square
fit to calculate the exponent.
Richeton {\it et al.}\ (2005) apply the Levenberg-Marquardt
algorithm for this purpose.
These methods are appropriate for fitting regression
curves, but they are not the most statistically efficient
techniques for a parameter evaluation in statistical
distribution.
They may yield biased estimates of an exponent parameter and
its uncertainty (Vere-Jones, 1988; Clauset {\it et al.},
2009).

In Section~\ref{analys} below we briefly review the
earthquake size distribution formulas and consider the results
of the $\beta$-value evaluation.
Section~\ref{syst} discusses systematic and random effects in
earthquake size determination (items 4-7 above).
In Section~\ref{disc} we summarize the results for determining
$\beta$ and compare our conclusions with studies of
dislocation avalanches in brittle and plastic crystalline
solids.
We also discuss possible consequences for interpreting
geophysical observations of regular and slow earthquakes and
seismic tremors in brittle crust and the upper mantle.

\section{Catalog analysis and earthquake size distribution }
\label{analys}

Only shallow earthquakes (depth 0-70~km) will be investigated
in this work, because more data is available on them.
Additionally, the seismic efficiency coefficient, or
proportion of tectonic deformation released by such
events, is close to 1.0 (Bird and Kagan, 2004).
Geometry of deep earthquake faults is much less known than
that for shallow seismicity, and may be more complex.
For deeper earth layers (depth $>$~70~km) only a small part
(less than 5\%) of tectonic motion is released by seismicity
(Kagan, 1999; Frohlich and Wetzel, 2007).
Therefore, earthquake rupture in a brittle crust would be
better modeled by the critical branching process theory.

As mentioned in the Introduction, regular magnitude
measurements are subject to many random and systematic errors
(Kagan, 2003).
Kagan (1999, pp.~557-8) studied correlation between $b$-value
estimates based on $m_b$ and $M_S$ magnitudes and found that
the correlation coefficient is low (0.1--0.2).
These coefficient values seem to indicate that $b$-value
variations are not caused by regional tectonic or physical
factors.
Therefore, we investigate earthquake size distribution here
using only seismic moment tensor measurements.
The most complete, extensive, and accurate catalog of tensor
solutions is the CMT dataset (Kagan, 2003).

\subsection {CMT earthquake catalog
}
\label{catl}

We studied earthquake distributions and clustering for the
global CMT catalog of moment tensor inversions compiled by
the CMT group (Ekstr\"om {\it et al.}, 2005; Ekstr\"om, 2007).
The present catalog contains more than 30,000 earthquake
entries for the period 1977/1/1 to 2008/12/31.
Earthquake size is characterized by a scalar seismic moment
$M$.
The moment magnitude can be calculated from the seismic moment
(measured in Nm) value as
\be
m_W \ = \ (2/3) \cdot \log_{10} M - 6.0
\, .
\label{Eq1}
\ee
The magnitude threshold for the 1977-2008 catalog is
$m_t=5.8$;
for the 1982-2008 catalog it is $m_t=5.6$ (Kagan, 2003).
An earthquake catalog is considered reasonably complete at and
above the magnitude threshold (or the corresponding seismic
moment, Eq.~\ref{Eq1}).
Since we use only the moment magnitude in this work, the
subscript is usually omitted.

\subsection{Seismic moment distribution }
\label{ana1}

The distribution of earthquake size is usually described by
the classical G-R (Gutenberg and Richter, 1944)
magnitude-frequency relation
\be
\log_{10} N (m) = a_t - b \, (m - m_t)
\qquad {\rm for} \quad m_t \le m,
\label{Eq2}
\ee
where $N(m)$ is the number of earthquakes with magnitude $\ge
m$, and $a_t$ and $b$ are parameters:
$a_t$ is the logarithm of the number of earthquakes with $m
\ge m_t$ (the seismic activity or earthquake productivity
level) and $b$ describes the relation between numbers of small
and large earthquakes.
The G-R laws seems to apply to earthquakes as small as
$m=-1.3$ (Boettcher {\it et al.}, 2009), with a rupture length
on the order of 0.5~m, and as large as the 2004 $m9.2$
Sumatra, with about a 1200~km rupture.

The G-R relation (\ref{Eq2}) can be transformed into the
Pareto (power-law) distribution for the scalar seismic moment
$M$.
The distribution in a probability density form is
(Kagan, 2002a)
\be
f (M) \ = \ \beta \, M_t^\beta \, M^{ -1 - \beta}
\qquad {\rm for} \quad M_t \le M \, ;
\label{Eq3}
\ee
in a survivor function ($1 \ -$ cumulative distribution)
form it is
\be
F (M) \ = \ \ (M_t/M)^{\beta} \,
\qquad {\rm for} \quad M_t \le M,
\label{Eq4}
\ee
where $\beta$ is the index parameter of the distribution, and
$b = {3 \over 2} \, \beta $.

The tapered G-R (TGR) relation has an exponential taper
applied to the number of events with a large seismic moment.
Its probability density function is
\be
f (M) \ = \ \left [ \, \beta + {M \over M_c} \, \right ] \,
M_t^\beta \, M^{ -1 - \beta}
\exp \left ( {{M_t \, - \, M} \over M_{c} } \right ) \quad
{\rm for} \quad M_t \le M < \infty \, .
\label{Eq5}
\ee
Here $M_{c}$ is the parameter controlling the distribution
in the upper ranges of $M$ (`the corner moment').
The survivor function is
\be
F (M) \ = \ (M_t/M)^{\beta} \,
\exp \left ( {{M_t \, - \, M} \over M_{c} } \right ) \quad
{\rm for} \quad M_t \le M < \infty \, .
\label{Eq6}
\ee
Equations (\ref{Eq5}) and (\ref{Eq6}) are equivalent
to (\ref{Eq3}) and (\ref{Eq4}), respectively, if $M_c \to
\infty$.

\begin{figure}
\begin{center}
\includegraphics[width=0.75\textwidth]{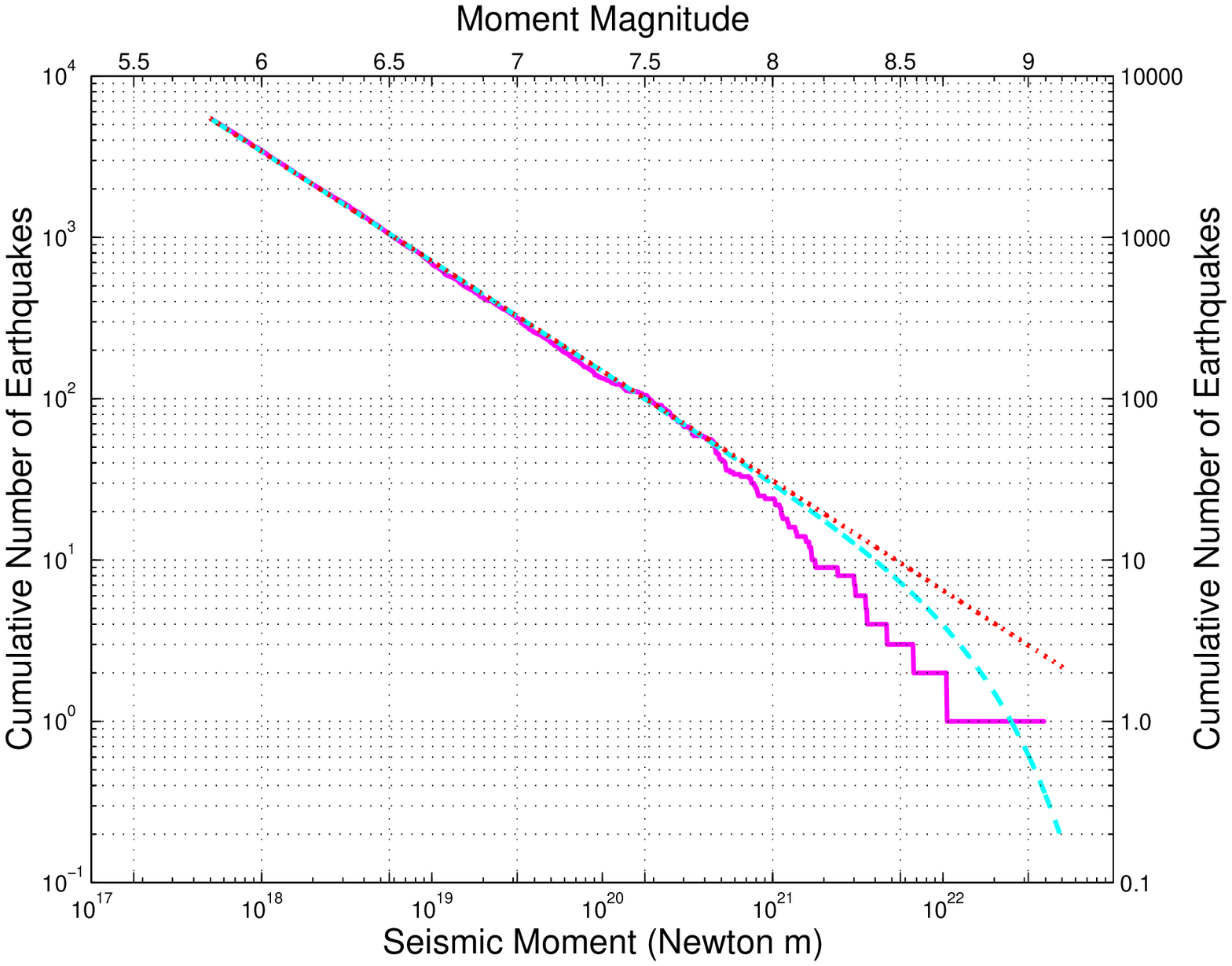}
\caption{\label{fig1}
}
\end{center}
Number of earthquakes with moment ($M$) larger than
or equal to $M$ as a function of $M$ for the shallow
earthquakes in the CMT catalog during 1977--2008,
moment threshold $M_t=10^{17.7}$~Nm ($m_t=5.8$), the total
number of events 5450.
Power-law approximation (equivalent to Gutenberg-Richter law)
is shown by dotted line.
Dashed line shows tapered Gutenberg-Richter distribution: the
G-R law restricted at large seismic moments by an exponential
taper with the corner magnitude $m_c=8.9$.
The slope of the linear part of the curve corresponds to
$\beta=0.68$.
\end{figure}

Fig.~\ref{fig1} displays cumulative distribution (survivor
function) for the scalar seismic moment of shallow earthquakes
in the CMT catalog for 1977-2008.
The curves display a scale-invariant (Pareto) segment (linear
in the log-log plot) for small and moderate values of the
moment magnitude $m$.
But for large $m$, the curve clearly bends downward
(see Fig.~2 by Kagan {\it et al.}, 2010 for comparison).

\begin{figure}
\begin{center}
%
%
\includegraphics[width=0.75\textwidth,angle=-90]{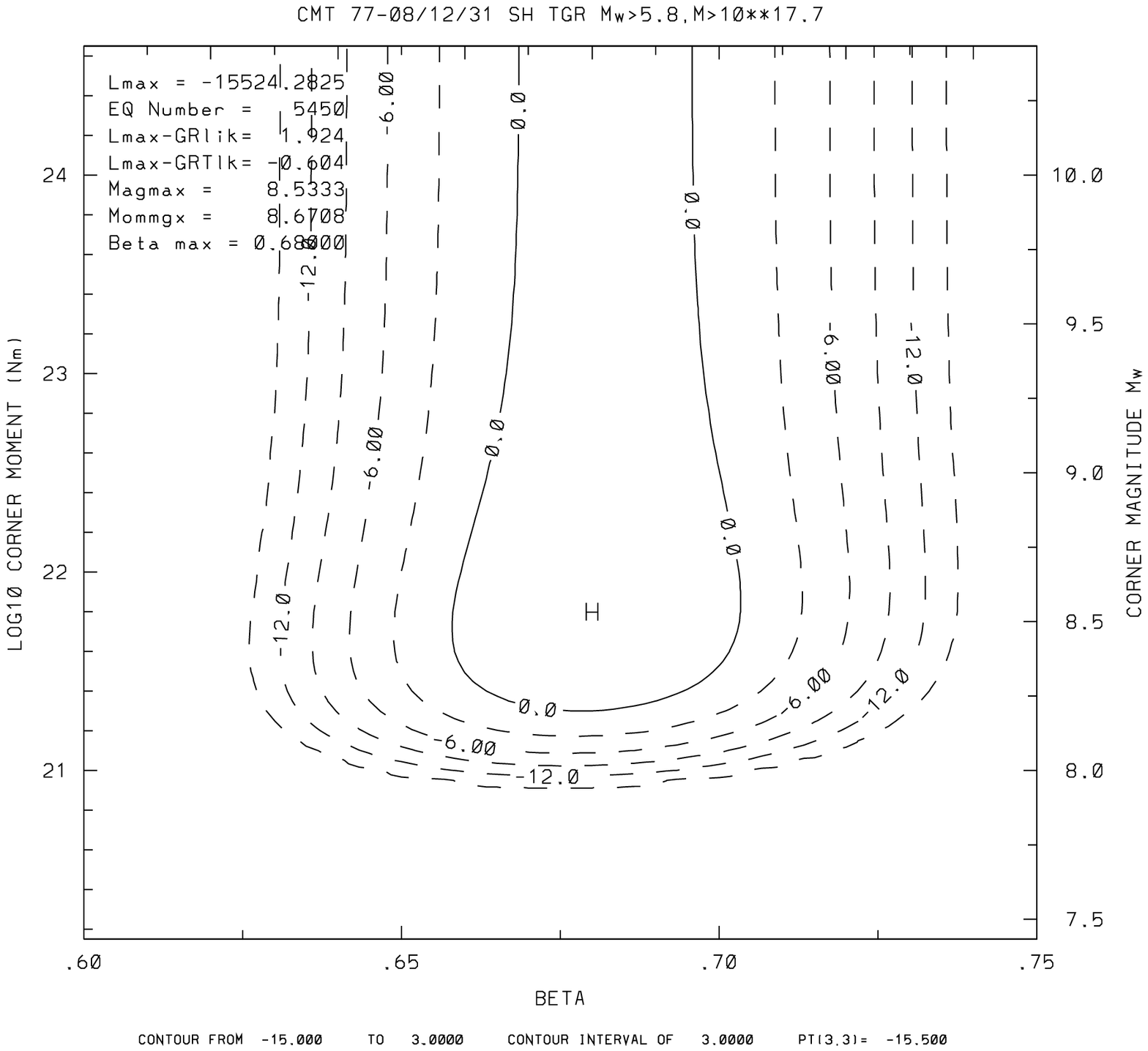}
\caption{\label{fig2}
}
\end{center}
Log-likelihood maps for the distribution of scalar seismic
moment of earthquakes:
The CMT catalog time span is 1977 January 1 -- 2008 December
31; the seismic moment cutoff is $10^{17.7}$ Nm ($m_t = 5.8$);
the number of events is 5450.
$H$-sign on the plot denotes the maximum likelihood estimate
of the parameters of interest.
\end{figure}

In a regular $\beta$ estimation we construct a likelihood map
in $[\beta \times m_c]$ space (Fig.~\ref{fig2}) and find its
maximum (compare Figs.~6 and 7 by Bird and Kagan, 2004, or
Fig.~3 by Kagan {\it et al.}, 2010).
The 95\% confidence level corresponds to the contour value 0.0.
Only the lower confidence limit for the corner magnitude
$m_c \simeq 8.3$ can be determined.
The upper limit is not defined in the map.
The tapered Pareto distribution with the parameters
corresponding to the maximum likelihood function, is shown in
Fig.~\ref{fig1} by the dashed line.
Sometimes these maps exhibit a significant correlation between
the estimates of parameters, see Figs.~6~G,H and 7~B,E by Bird
and Kagan (2004), complicating $M_c$ evaluation and
application.
Since our map does not display the interdependence between
these two parameter estimates, simpler procedures for
determining $\beta$ can be used (Kagan, 2002a).

\begin{figure}
\begin{center}
\includegraphics[width=0.75\textwidth]{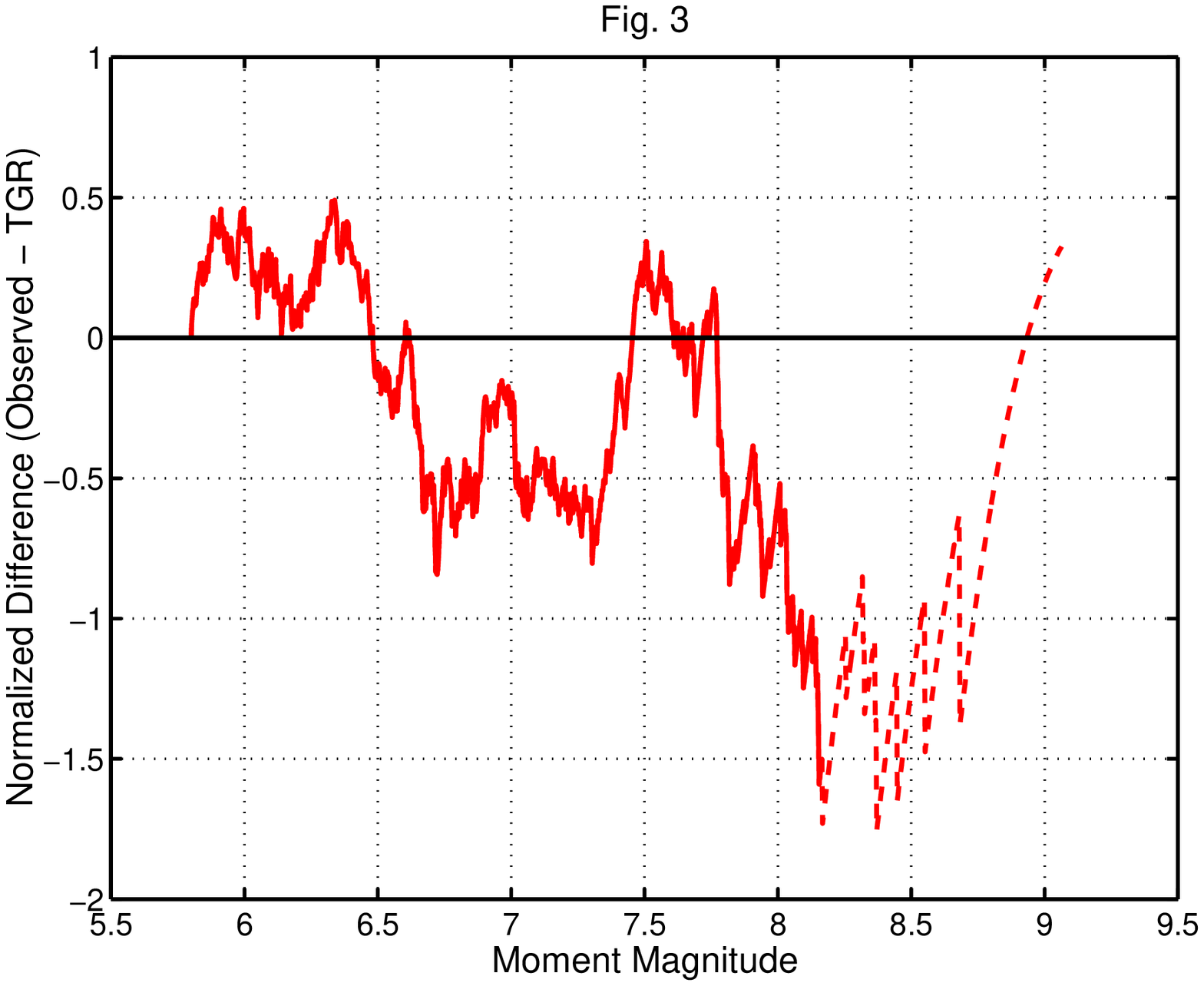}
\caption{\label{fig3}
}
\end{center}
Difference between observed magnitude-frequency relation and
its approximation by tapered G-R law (see Fig.~\ref{fig1}).
CMT catalog 1977--2008, magnitude threshold $m_t=5.8$.
The dashed line indicates where the difference is based on
fewer than 10 events.
\end{figure}

In Fig.~\ref{fig3} we show the normalized difference between
the observed magnitude-frequency relation in Fig.~\ref{fig1}
and its approximation by the tapered Pareto or the tapered
Gutenberg-Richter (TGR) relation.
We assume that the earthquake numbers in a frequency plot
follow the Poisson distribution.
Kagan (2010) demonstrates that the temporal distribution of
large earthquakes approaches the Poisson law.
Kagan {\it et al.}, (2010, Table~2) show that in the CMT
catalog due to its high moment threshold the aftershock
numbers are less than 25\% of the total.
For large number of events its standard deviation ($\sigma$)
is a square root of the number.
To normalize, we divide the difference by $1.96 \times
\sigma$; hence the $\pm 1$ ordinate value would correspond to
the 95\% confidence level.

\begin{figure}
\begin{center}
\includegraphics[width=0.75\textwidth]{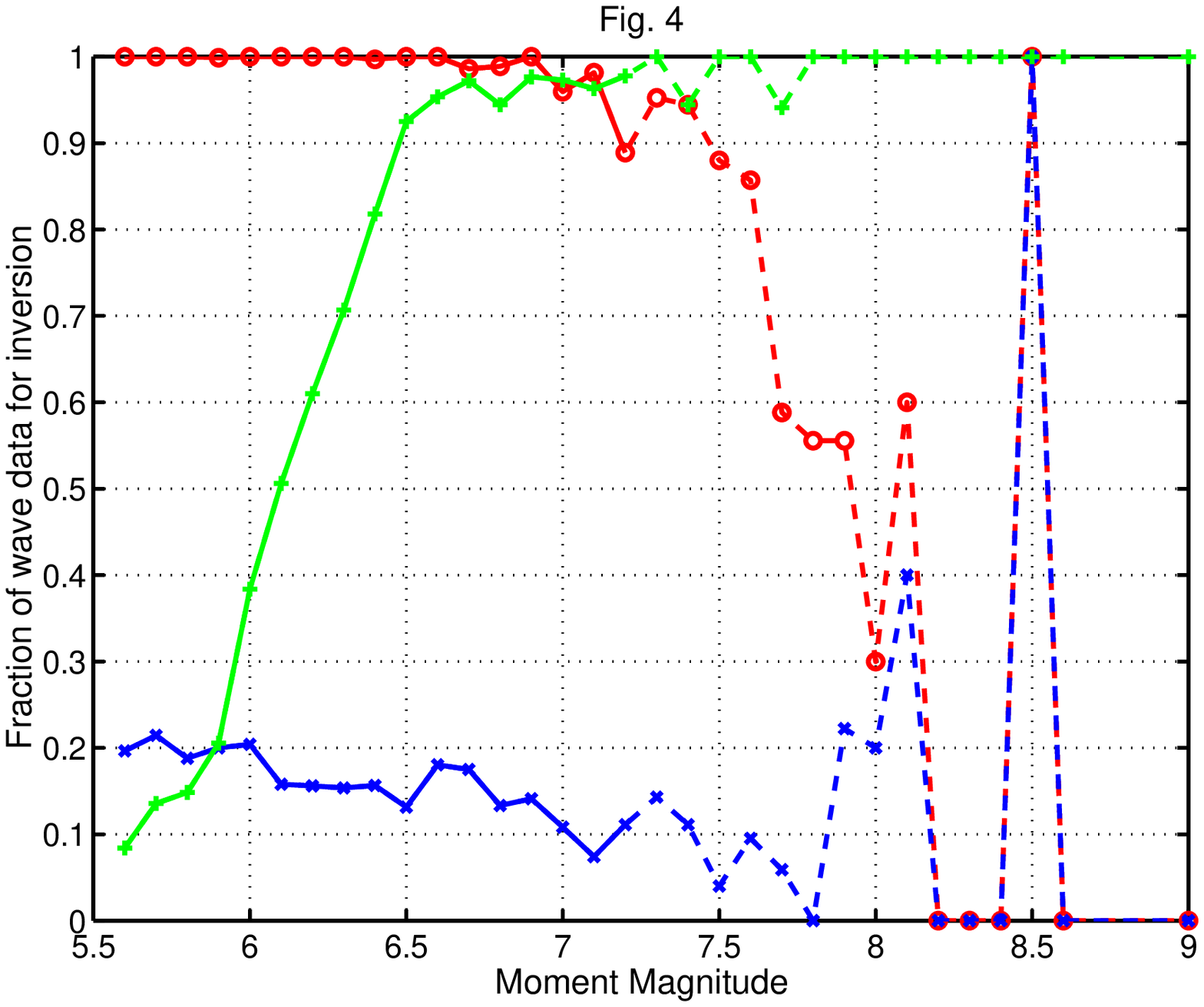}
\caption{\label{fig4}
}
\end{center}
Percentage of moment tensor solutions inversion of CMT data
based on different waves:
body waves -- red line, circles; surface waves -- blue
line, x-marks; mantle waves -- green line, pluses.
Dashed lines are for data with fewer than 10 earthquakes in
the 0.1 magnitude interval.
\end{figure}

Fig. 4 demonstrates a possible source of bias in
determining magnitude: earthquakes of different size have
their size estimated differently
The moment inversion in the CMT catalog is
carried out using some combination of three types of
waves: body, surface, and mantle waves (Dziewonski {\it et
al.}, 1981; Dziewonski and Woodhouse, 1983; Ekstr\"om, 2007).
The mantle low-frequency (period 135~s) waves are used mostly
for larger earthquakes.
Higher-frequency body and surface waves are used for more
moderate events (see also Fig.~2 by Kagan, 2003).
For shallow earthquakes, Fig.~\ref{fig4} shows the fraction of
each wave used in the solution for each moment magnitude.
Earthquake sizes are estimated by some combination of the
different wave types, so the sum of three points in any
given magnitude bin may exceed unity.
For example, up to $m=7.5$ almost 90\% of earthquakes have
body waves implemented in the solution.
After that the inversion is based mostly on mantle waves.
Therefore, we may posit that the statistically significant
deviation for the tapered Pareto distribution at $m > 7.8$
(see Fig.~\ref{fig3}) may be caused by transition from the
body- to mantle-wave estimate.
It is possible, of course, that the change in both plots
(Figs.~\ref{fig3} and \ref{fig4}) at about $m=7.5$ is a
coincidence.
Presently we lack data to thoroughly test this hypothesis.

\section{Systematic and random effects in
determining earthquake size
}
\label{syst}

This section considers in detail four sources of
magnitude/moment bias mentioned in the Introduction (items
4-7).

\subsection{Scalar seismic moment errors }
\label{ana2}

There is a bias in evaluating the G-R parameters due to
random errors in determining magnitude (Molchan and
Podgaetskaya, 1973, p.~47, Eq.~7; Tinti and Mulargia, 1985,
p.~1690).
This magnitude error is assumed to be symmetric and Gaussian,
the error analysis by Kagan (2003, see Figs. 12-15) seems to
confirm this.
Generally, as long as the moment or amplitude errors are
relatively small, one should expect them to be Gaussian as
they are the result of summing up many independent random
uncertainties.

The magnitude error causes a shift of the $a$ estimate in
(\ref{Eq2}) toward larger values: given the approximate
symmetry of the error distribution, more weak earthquakes have
their magnitude increased than vice versa:
\be
a_{\rm {estim}} \ = \ a_{\rm {corr}} \ + \ {
{b^2 \, \sigma^2_m \, \log 10 } \over { 2 } } \, ,
\label{Eq7}
\ee
where
$\sigma_m$ is a standard error in the magnitude estimate,
$a_{\rm corr}$ is the corrected (true) $a$-value,
$a_{\rm estim}$ is the measured $a$-value.
Otherwise, the bias may manifest as a right-hand (horizontal)
magnitude shift in the G-R curve
\be
m_{\rm {corr}} \ = \ m_{\rm {estim}} \ - \ {
{b \, \sigma^2_m \, \log 10 } \over { 2 } } \, ,
\label{Eq8}
\ee
where
$m_{\rm corr}$ is the corrected magnitude,
$m_{\rm estim}$ is the measured magnitude.
For $b=1$ the shift is of the same amplitude in (\ref{Eq7})
and (\ref{Eq8}).

If the magnitude errors do not depend on the magnitude, this
error does not practically influence the estimated
$b$-value (Tinti and Mulargia, 1985).
However, if $\sigma_m$ is a function of $m$, the $b$
estimates would be affected by magnitude errors.

Rhoades (1996, 1997) derived the theoretical estimates of a
bias in the $b$-value in such a case.
Rhoades (1996) and Rhoades and Dowrick (2000) studied the
influence of magnitude errors on $b$-value estimates and
provided some approximate estimates for the $b$-bias from
such errors.
In these evaluations they assumed that magnitude
errors increase as the magnitude itself increases: for the
$j$-th measurement of magnitude $\sigma_j \ = \ 0.1 \, ( \, 1
+ u_j \, m_j \, ) $, where $u_j$ is a random number uniformly
distributed on the interval [0, 1], the magnitude threshold is
3.95 and the real $b$-value is 1.0.
They found that the `measured' $b$-value decreases by about
4\% compared to its true value.

The CMT catalog supplies inversion errors for tensor moment
components (Dziewonski {\it et al.}, 1981;
Dziewonski and Woodhouse, 1983).
It is important to distinguish magnitude and moment random
errors.
Magnitude is defined as a logarithm of appropriately scaled
and averaged amplitude ($A$) of specific seismic waves: $ m =
\log_{10} A + C$, where $C$ is a coefficient.
Therefore, for small amplitude uncertainties the standard
error in the amplitude measurements ($\sigma_A$) is related to
the magnitude error ($\sigma_m$) as
\be
\sigma_m \ \propto \ { {\partial \, m } \over {\partial A } }
\ = \ { \sigma_A \over A} \times C
\, ,
\label{Eq8a}
\ee
i.e., the magnitude error is proportional to a {\sl relative}
amplitude error.
A similar relation is valid for the scalar seismic moment (see
Eqs.~\ref{Eq9} and \ref{Eq12} below).

\begin{figure}
\begin{center}
\includegraphics[width=0.75\textwidth]{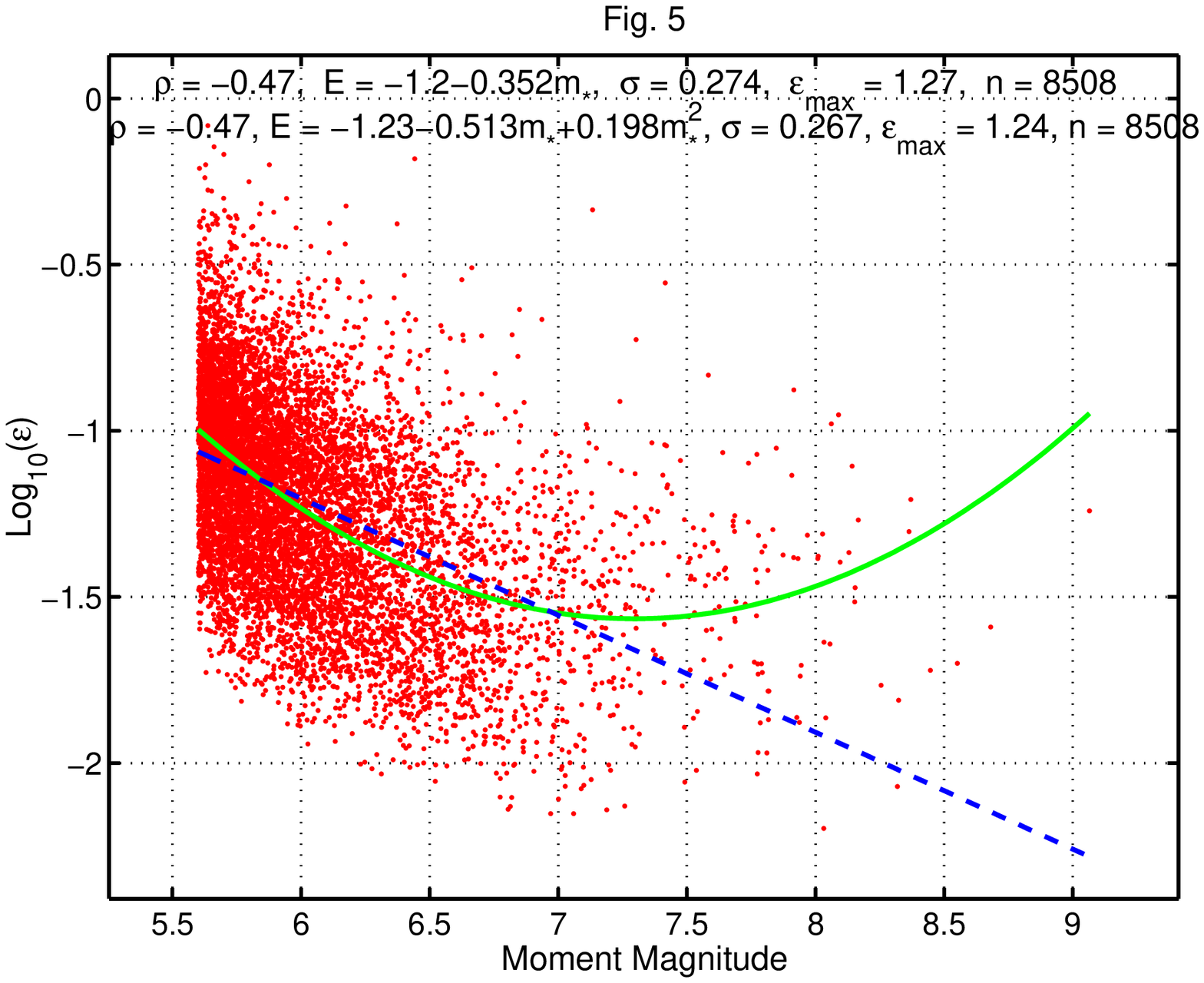}
\caption{\label{fig5}
}
\end{center}
Relative error $\epsilon$ versus moment magnitude for shallow
earthquakes $m_t = 5.6$ in the 1977--2008 CMT
catalog.
The curves show two approximations: linear and quadratic fits.
\end{figure}

Kagan (2002a) measured the scalar seismic moment errors in the
CMT catalog and found that the relative moment errors actually
{\sl decrease} with the earthquake size, implying that the
{\sl magnitude} errors should also decrease.
We define the relative moment error, $\epsilon$, as the
ratio of the error tensor norm to the moment tensor norm
\be
\epsilon = \sqrt { \, \sum_{i, j} E_{ij}^2 \Big / \sum_{i,
j} M_{ij}^2 \, } \, ,
\label{Eq9}
\ee
where $E_{ij}$ and $M_{ij}$ are standard error and moment
tensor components, respectively.
The distribution of $\epsilon$ for the CMT catalog 1977-2008
is shown in Fig.~\ref{fig5} (compare Fig.~5 in Kagan,
2002a).
Since the influence of the magnitude threshold value is
insignificant for determining the relative moment error, we
use $m_t = 5.6$ in this plot.
We calculate two regression lines approximating the dependence
of the errors on the magnitude: the linear and quadratic
curves
\be
\log_{10} \epsilon \ = \ c_0 \ + \ c_1 \, (m-6) \ + \ c_2
\, (m-6)^2 \, .
\label{Eq10}
\ee
We use $(m-6)$ instead of $m$ as an argument, so that the
$c_0$-value would have a clear intuitive meaning.

In the diagram (Fig.~\ref{fig5}) the coefficient of
correlation between $\epsilon$ and the magnitude is $-0.47$,
indicating that relative moment errors decrease with the
increase of $m$.
Residual regression errors are close for both linear and
quadratic cases: $\sigma = 0.274$ and $\sigma = 0.267$,
respectively.
From the diagram it is clear that errors for earthquakes with
$m > 6.5$ deviate significantly from a linear trend.
However, since the number of strong earthquakes is small, the
residuals of the linear and quadratic cases do not differ
significantly.

The parameter values in (\ref{Eq10}) for two subsets of the
CMT catalog are listed in Table~\ref{Table1}.
The $\epsilon$-values for earthquakes in the magnitude
range 5.4--6.4 can be represented as
\be
\epsilon \ \approx \ 0.056 \times 10^{- 0.54 \, (m - 6) }
\, ,
\label{Eq11}
\ee
for shallow events in the 1982-2008 CMT catalog (see
Table~\ref{Table1}).
For small $\epsilon$ the magnitude error $\sigma_m$ is
calculated as
\be
\sigma_m \ \approx \ { \epsilon \over { 1.5 \, \log 10 } }
\, .
\label{Eq12}
\ee
Modifying (\ref{Eq8})
or Eq.~10 in Rhoades and Dowrick (2000), we obtain the
following magnitude correction for the magnitude estimates
perturbed by random errors:
\be
m_{\rm corr} \ = \ m_{\rm estim} \ - \ {3 \over 4} \,
\sigma_m^2 \, \beta \, \log 10 \, .
\label{Eq13}
\ee

To apply this formula to $b$ or $\beta$ correction, we
estimate $\sigma_m$ at two magnitude values (5.4 and 6.4) and
use (\ref{Eq13}) to compute $\delta_c m = m_{\rm corr} -
m_{\rm estim}$.
Performing such calculations for the relative moment error
$\epsilon$ in (\ref{Eq12}), we obtain the correction for
$\beta$ of shallow earthquakes 0.0013: a $\beta$ is decreased
by about 0.2\%.

However, our calculations could not consider one
important source of error.
For many weak shallow earthquakes in the CMT catalog,
no solution can be obtained for the tensor components
$M_{r\theta}$ and $M_{r\phi}$ (Dziewonski {\it et al.}, 1981,
p.~2829; Dziewonski and Woodhouse, 1983; Frohlich and Davis,
1999).
In such a case, $E_{r\theta}$ and $E_{r\phi}$ as
well as $M_{r\theta}$ and $M_{r\phi}$ are set to zero.
About 4\% of shallow earthquakes have this problem (Kagan,
2003, pp.~195-196).
For strike-slip events which predominate in this group,
the tensor components $M_{r\theta}$ and $M_{r\phi}$ are close
to zero.
This means that if the values of $E_{ij}$ and $M_{ij}$ were
available for these events in (\ref{Eq9}), the numerator value
should be much greater, but the denominator would be
essentially the same.
This would significantly increase the resulting
$\epsilon$-value.

Moreover, apparently the relative moment error, $\epsilon$, is
only part of the total seismic moment error.
Dziewonski {\it et al.}\ (1981) and Dziewonski and Woodhouse
(1983) suggested the standard errors obtained in the CMT
solutions likely to be underestimated.
Kagan (2000, 2002a) estimated that the reported errors are
possibly 1/3 to 1/2 of the total.
Because the bias in estimating $\beta$ depends on the square
of the magnitude estimation error (see Eq.~\ref{Eq13}), a
systematic bias as high as 1-3\% may be caused by the decrease
of the relative magnitude uncertainty with the increasing
earthquake size.

\subsection{Earthquake sequences and their influence }
\label{ana3}

As mentioned in the Introduction, theoretical $\beta$
estimates should be only relevant for earthquake {\sl
sequences}, not individual events.
We take that an earthquake belongs to sequences and sequences
being the theoretical entity of interest, rather than
individual earthquakes.
Registration of aftershock sequences (Kagan, 2004; Enescu
{\it et al.}, 2009) shows that immediate aftershocks
observed in high-frequency seismograms are included in
mainshock or large aftershocks in catalogs based on long
period registration.
Thus, for example, the CMT catalog earthquakes include some
close aftershocks.

In our model earthquake sequences are produced by the same
critical branching process.
Later aftershocks are separated into individual events due to
temporal delays controlled by Omori's law (Kagan and Knopoff,
1981).
Occasionally, the first event in a sequence is weaker than the
following events, in which case it is commonly called a
`foreshock' ({\it ibid}).
Therefore, we could consider an earthquake cascade or
foreshock-mainshock-aftershock sequence as one entity.

Here we attempt to study the seismic moment distribution for
earthquake sequences.
To define the sequences, we use the results of the likelihood
analysis of earthquake catalogs (Kagan, 1991b).
We approximate an earthquake occurrence by a multidimensional
Poisson branching cluster process.
In this model spontaneous events are distributed according to
the Poisson distribution, whereas dependent events in
earthquake sequences are controlled by a distribution
characterized by a few adjustable parameters.
Each event, even if it is a member of a cluster, may start its
own sequence, hence aftershocks may be of the first-, second-,
third-, etc., order.
The parameters of the model are estimated through a maximum
likelihood search.
A similar scheme has been proposed recently by Zhuang {\it et
al.}\ (2002, 2004).

As the result of the likelihood optimization, we evaluate
probabilities ($p_{ij}$) of any $i$-th earthquake belonging to
a $j$-th cluster or sequence ($\sum_j p_{ij} = 1$); $p_{ii}$
corresponds to the probability that an earthquake will be
considered independent.
We use these probabilities ($p_{ij}$) to assign a part of the
seismic moment of the $i$-th event to the $j$-th earthquake;
the $j$-th earthquake might again belong to some $k$-th group,
etc.
This process continues until all earthquakes and their
interconnections in a catalog have been counted.
In the end, some of the aftershock moments are transferred to
their mainshocks if the aftershock probably belongs to the
particular mainshock's cluster.
As a result of this seismic moment reassignment, some
earthquakes may have a seismic moment below the magnitude
threshold, $m_t$.
We remove these earthquakes from a catalog.

As the number of earthquake sequences is always smaller than
the number of earthquakes in a catalog, while the total
moment in a catalog is constant, we should expect the
$\beta$-value for sequences to be smaller than that
for individual earthquakes.
For deep and intermediate earthquakes, the difference in the
$\beta$-values in calculations which use sequences and those
using individual earthquakes is negligible.
This small difference is due to a small number of aftershocks
for these sequences in the CMT catalog (Kagan, 1999, Table~4).

Generally, we can treat the probabilities of being independent
($p_{ii}$) as corresponding to the weight of an earthquake as
it is included in calculations.
However, to make our computations similar to those used for
real catalogs, we simulate new catalogs leaving in only
earthquakes whose $p_{ii}$ exceeds a random number
distributed uniformly in the [0-1] interval.
Thus, we obtain a `declustered' catalog, in which we delete an
earthquake according to its probability of being dependent.

Table~\ref{Table2} shows several $\beta$ measurements for two
CMT shallow earthquake subcatalogs, 1977-2008 and 1982-2008,
with the magnitude threshold $m_t=5.8$ and $m_t=5.6$,
respectively.
For global datasets three types of computation were performed:
(a) in the original list, (b) in a declustered catalog,
where seismic moment has been preserved for each
earthquake, and (c) in a declustered catalog with aftershock
moment transferred to an appropriate mainshock according to
probabilities $p_{ij}$ (see above).
We performed similar measurements for earthquakes in
subduction zones (trenches) (Kagan {\it et al.}, 2010).
Trench earthquakes have not been declustered, because some
may have connections to outside events.
Therefore the dependence probabilities can be biased.

In Table~\ref{Table2} the $\beta$-values are smaller by about
1-3\% for the 1982-2008 dataset compared to the 1977-2008
catalog.
Probably a higher average accuracy of these solutions (Kagan,
2003) and larger magnitude range explains this reduction.
Bird and Kagan (2004) showed that for global seismicity the
minimum value of $m_c$ for some tectonic zones is of the order
5.9--6.6.
Extending the magnitude threshold to $m_t=5.6$ expands the
power-law part of the plot, and the influence of the corner
magnitude is smaller.

The $\beta$ bias in the Table is also caused by the mix of
different earthquake populations with various corner
magnitudes, $m_c$ (item~3 in the Introduction).
This effect could explain why the $\beta$-values for trenches
are significantly lower (by about 5\%) than the global ones.
A global earthquake set consists of many populations, of which
oceanic rift zones have the smallest $m_c$-values (Bird and
Kagan, 2004).
These oceanic events are excluded from the subduction zone
(trench) dataset.
Hence, the estimate of $\beta$ for trench earthquakes is
closer to the theoretical value than for earthquakes in any
other tectonic province.

As expected, the $\beta$-values decrease for the declustered
catalogs, since excluded aftershocks have smaller moment
values.
This reduction is even stronger for catalogs where the
aftershock moment is assigned to their potential mainshocks.
The $\beta$-value decreases are about 4.5\% and 8.5\%,
respectively.

These bias estimates depend on the correctness of the
calculations used to estimate earthquake probabilities.
The likelihood procedure used to assign the probability
of event independence is influenced by catalog quality,
length and magnitude threshold.
Given the presence of temporal boundaries, many relations
between earthquakes are missed: some events of the beginning
of the catalog may be aftershocks of previous strong
quakes.
Thus, instead of having a probability closer to zero, as they
should, these aftershocks would have an independence
probability equal to 1.0.

Due to the magnitude threshold, some connections between
events are not observable.
Suppose there is a potential foreshock-mainshock pair: a larger
earthquake is preceded by a smaller one.
However, the first event is below the magnitude threshold and
the larger quake is above it.
Then this second event would be treated as independent;
our calculations would not include this connection (Kagan,
1991b).

Moreover, the likelihood model used in our inversion is not
perfect (Kagan, 1991b).
As a result, the independence probability values of
earthquakes may not be fully counted leading to a bias in the
$\beta$ computations.
Therefore, the reduction of the $\beta$-values due to the
influence of aftershock sequences is likely to be greater.

If we add up all the influences of aftershocks, we should see
the $\beta$-value decrease to about 0.59--0.6.
Kagan (1991a, p.~129) and Kagan (1999, Table~5) obtained a
similar result for declustered shallow earthquakes.

\subsection{Seismic moment tensor and its complexity }
\label{ana4}

The previous discussion assumed that scalar seismic
moment is a fair measure of earthquake size.
In reality seismograms are caused by excitation from many
subevents during the main phase of an earthquake.
Thus, the seismic moment tensor of an earthquake
is a compound tensor sum of subevents.
If all these subevents were identically oriented, the tensor
sum would be proportional to a scalar sum of all the subevent
scalar moments.
However, detailed studies of earthquakes clearly indicate that
subevent orientation changes significantly during
ruptures.

The Bulletin of the Seismological Society of
America (BSSA) published several special issues dedicated to
thorough analysis of several large earthquakes like the 1992
Landers, 1999 Hector Mine, the 1999 Chi-Chi, Taiwan, the 2002
Denali, the 2004 Sumatra, and so on.
These studies detail a very complex geometrical picture of the
quake rupture process.
The focal planes and slip vectors of earthquake subevents
often rotate several degrees and even tens of degrees.
Therefore, the seismic moment tensor solution and the
resulting estimate of an earthquake scalar moment is subject
to random fluctuations from the stochastic misalignment of
earthquake components.

In principle, we could avoid the systematic effect caused by
source complexity if we used an earthquake's energy as a
measure of its size.
Energy is a positive scalar; thus, no bias due to source
complexity would appear in the energy estimate.
Unfortunately, estimates of the radiated seismic energy are
highly uncertain and often differ by up to an order of
magnitude (Perez-Campos {\it et al.}, 2003).
In contrast, the relative accuracy of evaluating seismic
moment tensor is on the order of $10^{0.15}$ (Kagan, 2003).

Scalar moment earthquake estimates should always be lower than
the sum of the subevents' scalar moments.
This occurs because of random fluctuations during earthquake
fracture.
This effect would also bias upwards the estimated
$\beta$-values.
Because we lack a comprehensive model of the earthquake rupture
process which would enable us to estimate rigorously the
resulting bias, we proceed by studying several approximations.
These will give insight into the problem and provide an order
of the magnitude estimate of possible systematic effects due
to source complexity.

We can estimate the influence of source complexity on the
resulting estimate of earthquake size by initially assuming
that small elementary subevents have their sign selected
randomly.
If the sign changes with equal probability ($p=0.5$), the
resulting sum of the subevents would be an ordinary random
walk.
The walk would converge to Brownian motion if the number
of subevents, $n$, is large.
The sum would be distributed according to the Gaussian
distribution with the standard deviation proportional to
$\sqrt n$.
The final value of the sum would be proportional to its
standard error.
(In a count of the Brownian sum, we use an absolute value
of the final walk position.
Therefore, the total `moment' estimate is positive.)
Therefore, in a critical branching process in which
descendants are added with the random probability $p=0.5$, the
power-law index would increase by a factor of two:
$\beta = 1.0$.

If we change the probability value from 0.5 to a higher level,
$p = 0.5 + \delta$, this would produce Brownian motion
with a drift (Feller, 1966).
For small $n$ values, the walk behavior would resemble a
regular Brownian motion, and later the sum would have a steady
component $n \, \delta$.
Thus, its behavior would be similar to the cumulative number
increase with $p=1$.

\begin{figure}
\begin{center}
\includegraphics[width=0.75\textwidth]{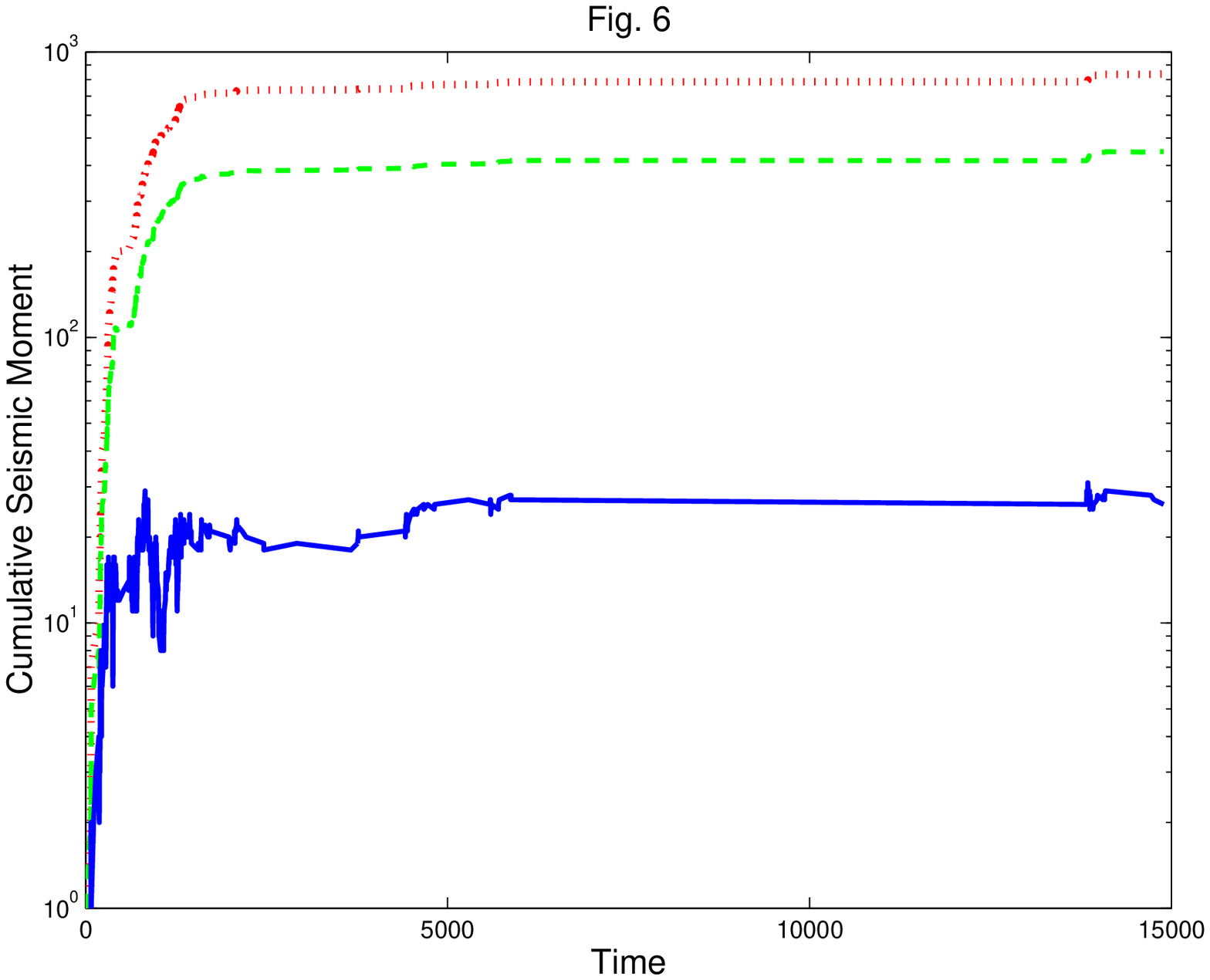}
\caption{\label{fig6}
}
\end{center}
Simulated source-time functions in a critical branching
process.
Red dotted line -- positive number addition ($p=1.0$).
Green dashed line -- unequal positive/negative number addition
(random walk with a drift, $p=0.75$).
Blue solid line -- equal positive/negative number addition
$p=0.5$ (random walk).
\end{figure}

Fig.~\ref{fig6} shows three simulated source-time functions
with the time delay controlled by the Omori-type function.
The cumulative functions for each curve are the sums of
elementary subevents of the unit size and only the signs are
different.
The first with $p=1$ (deterministic addition of events) is
similar to Fig.~3 by Kagan and Knopoff (1981).
The initial step-like increase of this function would likely
be interpreted as a mainshock, whereas a few steps at a later
time would be labeled as aftershocks.
The $p=0.75$ function increases the same way as the first
curve but with smaller amplitude; random fluctuations are not
easily observable.
The random walk function ($p=0.5$) behaves more erratically
and its total final ordinate is much smaller than that of the
other two curves.
Only the values of the curves at the end of a branching
simulation, corresponding to the total moment of a sequence,
are counted in our calculations.
For the $p$-values close to 0.5, the random branching walk
could end up as a negative cumulative sum; as we explained
above, we take the absolute value of the final ordinate of a
simulation run.
These values are assumed to correspond to the total seismic
moment of an aftershock sequence.

\begin{figure}
\begin{center}
\includegraphics[width=0.75\textwidth]{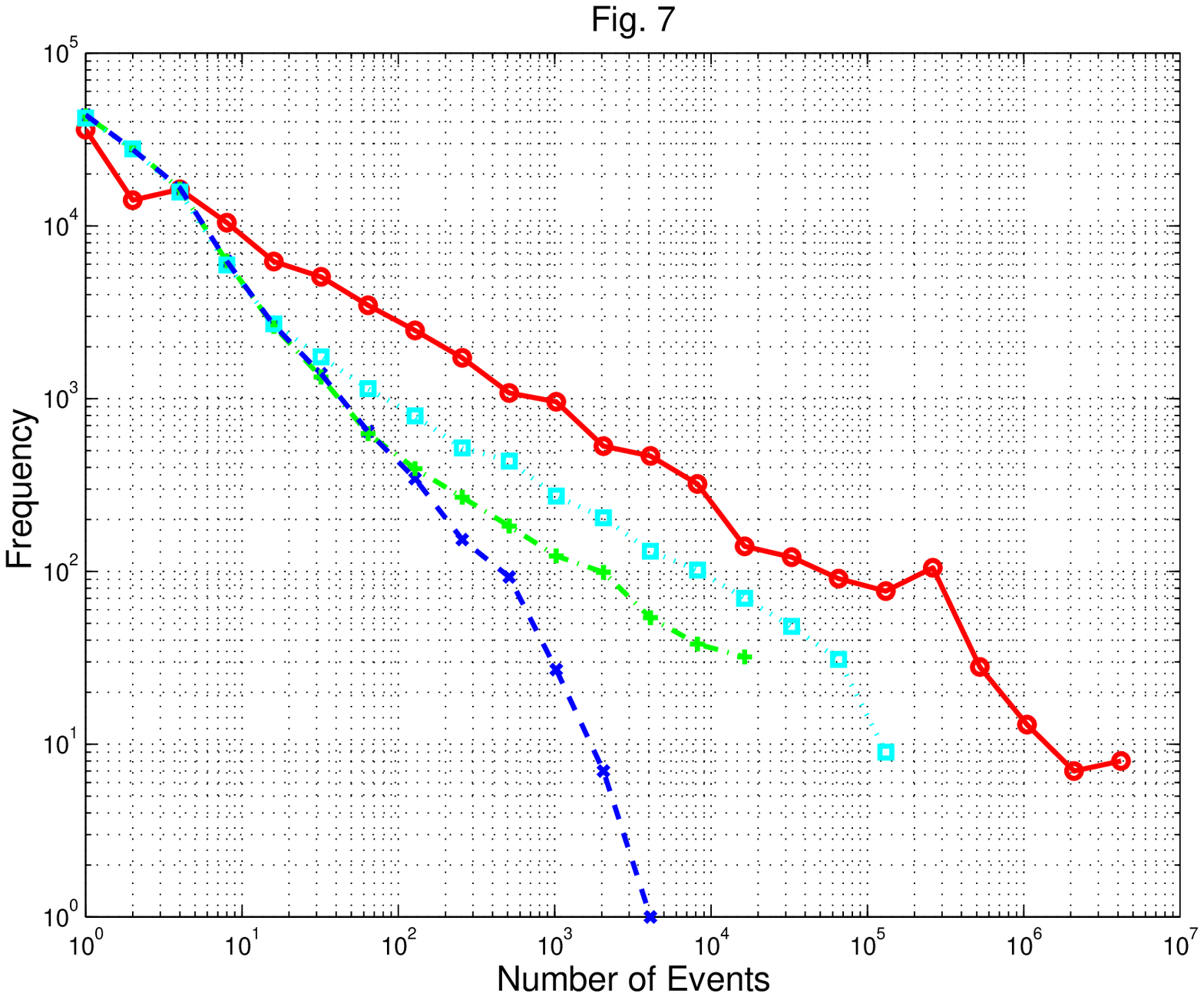}
\caption{\label{fig7}
}
\end{center}
Distribution of event numbers in a critical branching process.
Red line, circles -- positive deterministic number addition
($p=1.0$).
Blue line, x-marks -- equal positive/negative number addition
$p=0.5$ (random walk).
Green line, pluses -- unequal positive/negative number addition
(random walk with a drift, $p=0.51$).
Cyan lines, squares -- unequal positive/negative number
addition (random walk with a drift, $p=0.55$).
\end{figure}

Fig.~\ref{fig7} illustrates the above considerations.
We simulated a critical branching process and counted the
sum of events at the end of each simulation run, such as the
extreme right-hand points of three curves in Fig.~\ref{fig6}.
These numbers are shown in the diagram in the log-log format.
While the event numbers are small (less than 10), the
discretization effects are noticeable.
For the largest sequences, random fluctuations are observable,
because there are few of these sequences.
In the mid-number range, the deterministic number addition
($p=1$) distribution (red solid curve) has an index
$\beta=0.5$.
As expected, the Brownian walk addition (blue dashed curve)
has $\beta=1.0$.
As explained above, the curves for the motion with a drift
first follow the Brownian curve; then they are parallel to the
red curve.
Thus, in the beginning their index is $\beta=1.0$,
and for larger numbers it changes to $\beta=0.5$
confirming our predictions.
The randomness in the number addition significantly increases
the power-law distribution index.

\begin{figure}
\begin{center}
\includegraphics[width=0.75\textwidth]{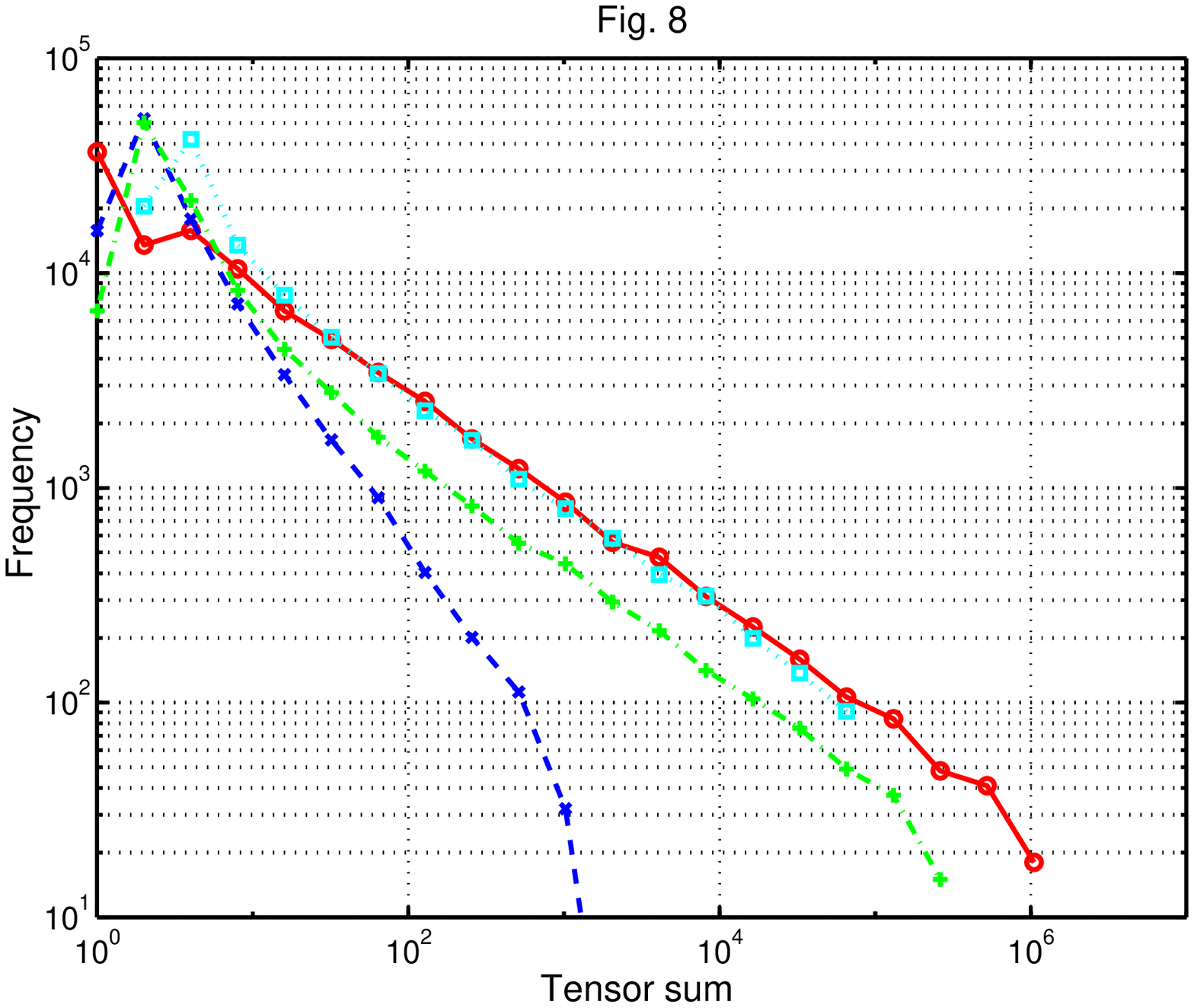}
\caption{\label{fig8}
}
\end{center}
Frequency plot of tensor sum norm in a critical branching
process.
Red line, circles -- tensor sum with no rotation,
the 3-D rotation angle, $\Phi=0^\circ$.
Blue dashed line -- random rotation ($\Phi=120^\circ$).
Green line, pluses -- limited random rotation,
$\Phi=80^\circ$.
Cyan line, squares -- limited random rotation,
$\Phi=30^\circ$.
\end{figure}

In Fig.~\ref{fig8} we show a more complicated test.
In a critical branching process simulation, we sum up seismic
moment tensors instead of scalar quantities.
In each simulation run, we determine a norm of the total
tensor sum which for the seismic moment tensor is equivalent
to the scalar moment.
Again the red solid curve shows the distribution when the
tensors are identical.
For other curves the tensors are independently randomly
rotated through the 3-D rotation angle, $\Phi$ (Kagan, 2003,
2009).
The maximum angle for a double-couple focal mechanism is
120$^\circ$; therefore, these tensors are rotated in a
uniformly random manner.
The angle $\Phi = 0^\circ$ corresponds to the zero rotation
(red curve).
If $\Phi < 120^\circ$, then the rotation is restricted, being
uniformly random only for angles smaller than $\Phi$.

The Fig.~\ref{fig8} diagram appears similar to the previous
plot (Fig.~\ref{fig7}).
If the tensors' orientation is identical, power-law exponent
$\beta=0.5$.
For a completely random orientation, $\beta=1.0$, and for a
restricted misalignment, the curves follow the latter
distribution first and are then parallel to the former line.

As discussed earlier, many earthquake ruptures exhibit
significant variations in focal mechanisms.
However, detailed analyses of individual earthquakes are
still rare and insufficient for rigorous statistical study.
Therefore, we study the degree of misalignment in several
mainshock/aftershock sequences.
Kagan (2000) investigated the correlation of earthquake focal
mechanisms and showed that the degree of mechanism 3-D
rotation increases between earthquakes with temporal and
spatial differences.
Hence, we hope that immediate aftershocks of strong
earthquakes will characterize the geometric complexity of
their rupture process.

To this end we studied all shallow (depth 0-70~km) earthquakes
in the 1977-2008 CMT catalog with a magnitude $m_1=7.5$ and
higher.
All earthquakes ($m \ge 5.6$) are considered aftershocks
within the first 7~days of
$m7.5$ earthquake occurrence and closer than
\be
r \ = \ 75 \times 10^{( \, m_1 \, - \, 7.5 \, ) \, /2} \ \
{\rm km} \, ,
\label{Eq14}
\ee
(Kagan, 2002b)
There are 105 mainshocks in the catalog and 81 of them
have one or more aftershocks.

To investigate the orientation differences between a mainshock
and its aftershocks, we calculate the correlation invariant
or tensor dot-product $J_3$ (Kagan, 1992; 2009)
\be
J_3 \ = \ \sum_{i, j} \ m_{ij} \, n_{ij}
\, ,
\label{Eq15}
\ee
for the main event ($m_{ij}$) and the sum of normalized
tensors for the whole 7-days aftershock sequence ($n_{ij}$).
Summation of repeating indices is assumed.
Both $m_{ij}$ and $n_{ij}$ are normalized.
In (\ref{Eq15}) $J_3 = 2.0$ means that focal mechanisms are
identical; $J_3 = -2.0$ corresponds to components of both
tensors having the opposite sign.

\begin{figure}
\begin{center}
\includegraphics[width=0.75\textwidth]{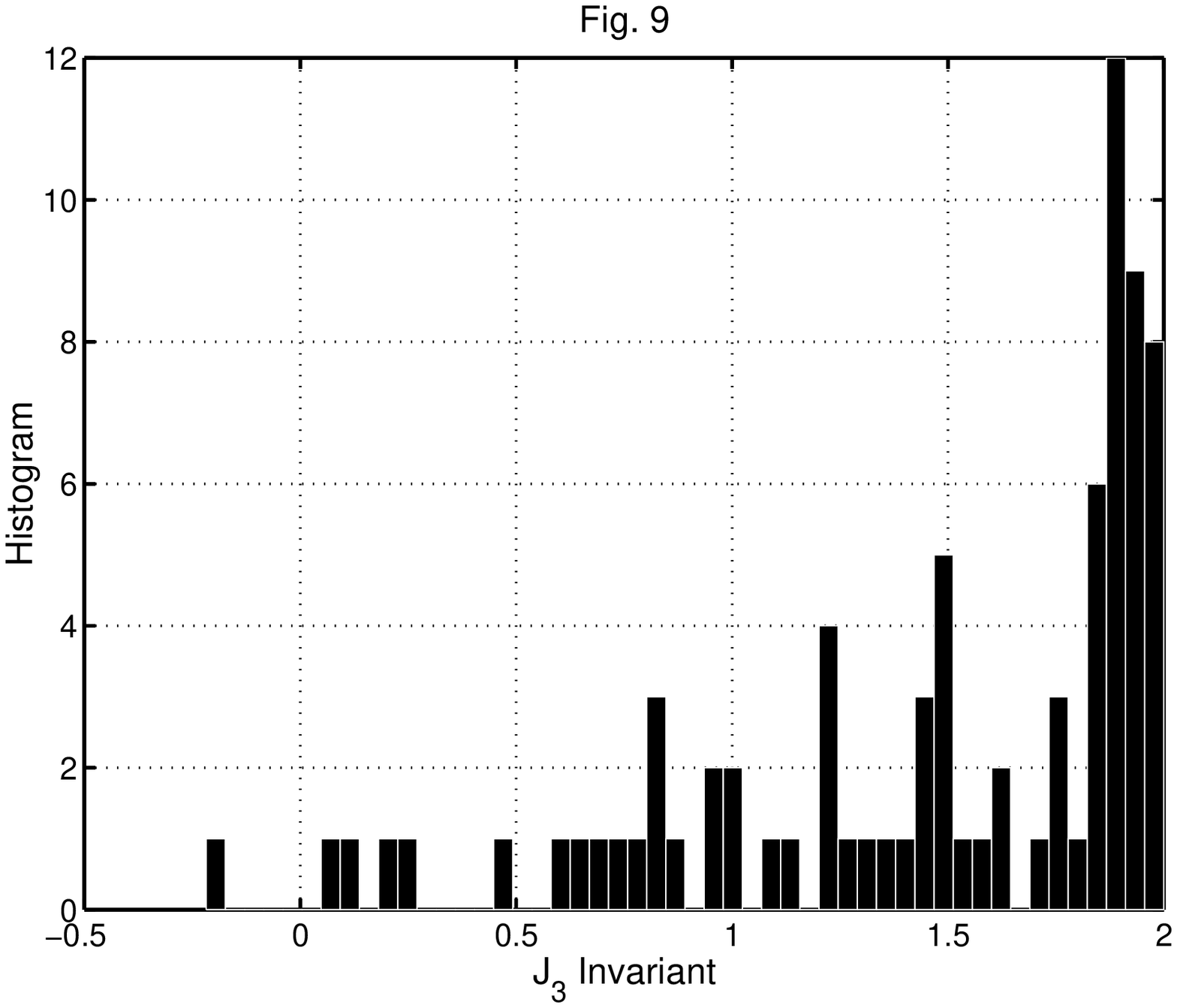}
\caption{\label{fig9}
}
\end{center}
Frequency plot of correlation tensor invariant (tensor
dot-product) for $m7.5$ mainshocks and sum of their immediate
aftershocks in 1977-2008 CMT catalog.
Average ${\overline {J_3}} = 1.458$, its standard error
$\sigma_J = 0.5535$.
\end{figure}

A $J_3$ histogram in Fig.~\ref{fig9} displays the correlation
between tensors.
Most correlation invariant values are close to 1.5--2.0.
Thus, the aftershock focal mechanisms are similar to that of
their mainshock.
However, some $J_3$-values are close to zero, and one
is negative, testifying to a significant variation in the
rupture process.
The smallest $J_3$-value is due to the November 2000 New
Ireland earthquake sequence.
The sequence started with a $m_W = 8.0$ left-lateral main
shock on 16 November and was followed by a series of
aftershocks with thrust mechanisms primarily (Geist and
Parsons, 2005; Park and Mori, 2007).
The negative $J_3$-value signifies that the aftershocks have
on average a slightly opposite orientation to their
mainshock.

\begin{figure}
\begin{center}
\includegraphics[width=0.75\textwidth]{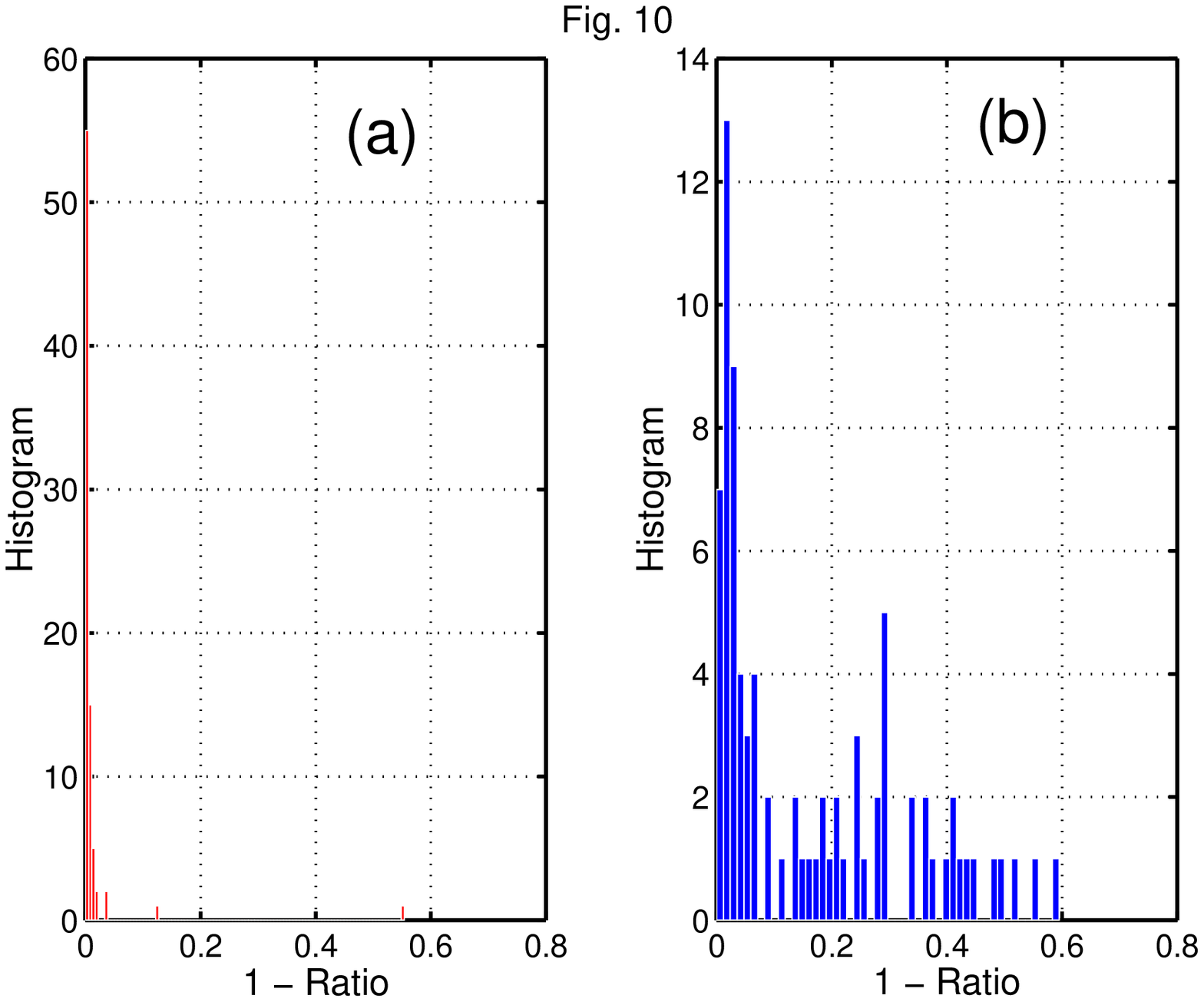}
\caption{\label{fig10}
}
\end{center}
Frequency plot of tensor/scalar sum ratio for $m7.5$ mainshocks
and immediate aftershocks in 1977-2008 CMT catalog:
(a) Unnormalized sum --
average ${\overline {(1 - R)}} = 0.0128$, its standard error
$\sigma = 0.0627$.
(b) Normalized sum --
average ${\overline {(1 - R)}} = 0.1642$, its standard error
$\sigma = 0.1639$.
\end{figure}

Fig.~\ref{fig10} displays two distributions of the ratio for
the tensor sum of the mainshock and its aftershocks to the
sum of their scalar moments
\be
R \ = \ | \sum \, m_{ij} \, | \, \, / \, \, \sum \ M \,
\, ,
\label{Eq16}
\ee
where $ | m_{ij} | $ means the norm of the tensor.
The aftershocks are selected according to the same criteria as
in Fig.~\ref{fig9}.
If aftershocks have the same focal mechanism as the mainshock,
the ratio would be 1.0.
In the left diagram of Fig.~\ref{fig10} the moment tensors are
not normalized; in the right plot they are.
In the inversions of the earthquake rupture process (see the
BSSA special issues mentioned above), several subevents of
approximately equal size but significantly different
orientation in focal mechanism are often observed.
This is the reason we investigate the normalized sums.

The diagrams show significantly varied focal mechanisms
in aftershock sequences.
Large fluctuations are seen in the normalized sums especially.
This result suggests again a conspicuous randomness occurs in
the focal mechanism orientation of earthquake sequences.
By implication this should also occur during an earthquake
rupture process.
Such random fluctuations may noticeably decrease the measured
earthquake size and influence the $\beta$ measurement.

What is the size of the $\beta$ measurement bias?
All estimates shown above are indirect.
Earthquake do not consist of an identically oriented or
purely random collection of elementary sources.
Various observations suggest that a rupture occurs over
quasi-planar fault patches, so there will be a strong
correlation between neighboring fault segments.
This correlation is sometimes broken by significant fault
branching.
Kagan (1982) proposed a geometrical model of such stochastic
rupture.
Unfortunately, the degree of geometrical branching ($\phi_0$)
in this model is not well known for different tectonic
provinces.
Therefore, we cannot easily simulate and study such branching
sequences.

What $\beta$-value change can be proposed as randomness result
in the fault rupture orientation?
Pure randomness yields $\beta$-value increase by a factor of
two (see Figs.~\ref{fig7}~and~\ref{fig8}).
Unfortunately, we cannot yet quantitatively study the
complex geometry of earthquake rupture.
We need to extrapolate from the measured misalignments of
close aftershocks.
These measurements indicate that complexity, though far from
being completely random, is nevertheless quite significant.
For example, in Fig.~\ref{fig9} the correlation invariant is
$J_3 = 1.46 \pm 0.55$.
These values can be compared to purely random arrangements of
double-couple focal mechanisms (Kagan, 1992), where
$J_3 = 0 \pm 0.89$ has been obtained.

\begin{figure}
\begin{center}
\includegraphics[width=0.75\textwidth]{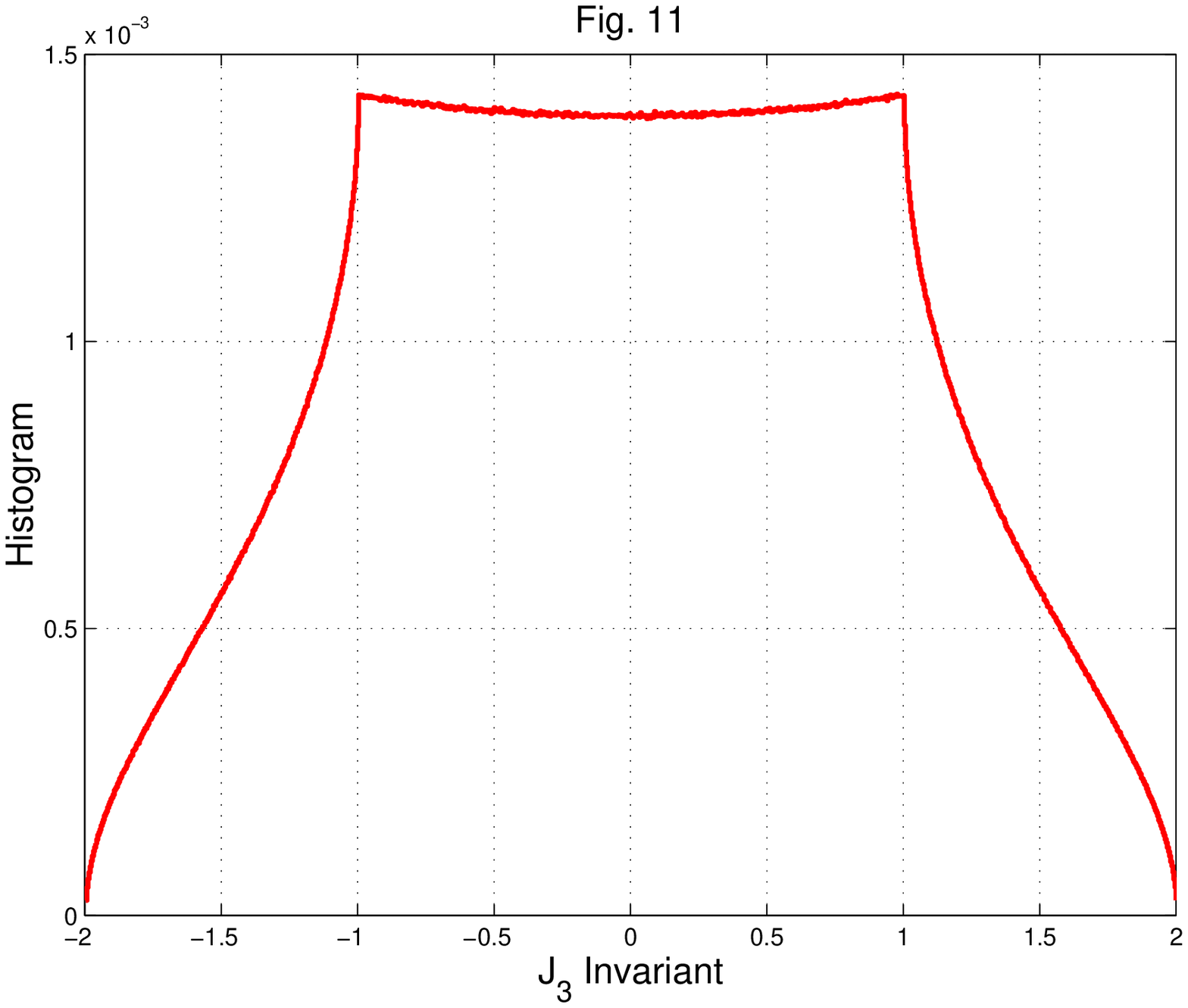}
\caption{\label{fig11}
}
\end{center}
Frequency plot of tensor dot-product invariant for random
rotation of double-couple sources.
Average ${\overline {J_3} } = 0$, its standard error $\sigma_J
= 0.8945$.
\end{figure}

The frequency plot for randomly oriented double-couples
obtained by simulation is shown in Fig.~\ref{fig11}.
As expected, the histogram curve is symmetric around $J_3=0$
and reaches the maximum at ${\rm abs}(J_3)=1$.
It would be interesting to obtain an analytical solution for
$J_3$, as we did for the 3-D rotation angle (Kagan, 2009).
This can be done in the future.

If focal mechanisms are all parallel, $J_3 = 2.0 $ (see more
in Kagan, 2009).
Therefore, we can make a rough guess that the degree of the
$\beta$-value increase would be on the order of 10-15\%,
when extrapolated to a time close to zero: during the
mainshock rupture.
This guess is obtained by comparing the average $J_3$ shift in
Fig.~\ref{fig9} (${\overline {J_3}} - 2.0 = 0.542$) with that
for the completely random arrangement in Fig.~\ref{fig11}
(${\overline {J_3}} - 2.0 = 0.0$),
and by comparing standard deviations for both
cases: 0.55 and 0.89, respectively.

Similar conclusions could be inferred from the results of the
3-D rotation angle distribution.
The average angle between the mainshock focal mechanism and
mechanisms of immediate aftershocks is on the order of
$10^\circ$ (Kagan, 2000).
For a completely random rotation, the maximum angle is
$120^\circ$ and the average angle is $75.2^\circ$ (Kagan,
2003).
Given the source complexity, as demonstrated by the rotation
angles, the $\beta$ bias should be around 8-12\%.

Analogous conclusions could be drawn from Fig.~\ref{fig10}.
If immediate aftershocks had the same moment tensor
solutions as their mainshock, the tensor/scalar sum ratio
$R$ in (\ref{Eq16}) should be 1.0, and $(1-R)$ equal zero.
Both average and standard deviation for $(1-R)$ in the
plots display significant non-zero values.
We infer that the source is complex and the $\beta$ bias may
be on the order of few percent.

To summarize the results of this subsection, we hypothesize
that as a consequence of the random geometrical misalignment
of a fault rupture, the measured $\beta$-value may be
increased by at least a few percent (3-7\%) from its true
size.
The estimate above is conservative.
More work needs to
be done to obtain a more reliable value which could
lead to an even greater $\beta$-estimate decrease.

\subsection{Centroid depth influence }
\label{ana5}

The CMT catalog supplies earthquake coordinates for the
seismic moment centroid (Dziewonski {\it et al.}, 1981; 1983);
the centroid is in the center of the moment release volume.
The centroid distance from the fault edge cannot be smaller
than a half of the earthquake rupture zone width.
Closer to the surface or to the fault boundary the corner
moment would approach zero.
An inspection of seismic maps suggests that hypocenters of
larger events are on average deeper than those for small
earthquakes.
Thus, the moment-frequency law (see Eqs.~\ref{Eq5} and
\ref{Eq6}) would change due to an increase of the maximum
earthquake size with depth (Kagan, 2002a, p.~539).
Therefore the corner moment for deeper earthquakes would
increase.

As we explained in the Introduction, the depth accuracy for
shallow earthquakes is presently insufficient to investigate
observationally the dependence of the corner moment on depth.
Therefore, we study a possible influence of the finite fault
size by calculating a new distribution of earthquake size for
a few simple models of the earthquake rupture pattern.
These theoretical guesses would help evaluate the depth
effect up to the order of magnitude.

We assume that earthquakes are distributed over an infinite
planar fault surface extending either vertically for 20~km
(imitating conditions for strike-slip faults in California),
or distributed over an inclined fault with a width of 200~km
(as in some subduction zones, see, for instance, Bird and
Kagan, 2004).
The variable corner moment for such faults is
\be
M_c^\prime \ = \ C \, \zeta^3
\, ,
\label{Eq17}
\ee
where $\zeta$ is distance from a centroid to a fault edge
and $C$ is an appropriate coefficient.

In one model we assume that earthquake centroids are
distributed uniformly over the fault surface.
Then, using the algebraic and numerical facilities of {\scaps
mathematica} (Wolfram, 1999), we calculate the new survivor
function
\be
\Psi_1 (M) \ \propto \ \int\limits_0^L F(M) \, d \zeta \ = \
\left ( M_t \over M \right )^{\beta} \, \int\limits_0^L \,
\exp \left ( {{M_t \, - \, M} \over {C \, \zeta^3} } \right )
\, d \zeta \ = \
\left ( M_t \over M \right )^{\beta} \,
{ { \Gamma \left ( - { 1 \over 3 }, \,
{ {M - M_t} \over {C \, L^3 } }
\right ) }
\over { 3 \left ( { C \over {M-M_t} } \right ) ^{1/3} } }
\, ,
\label{Eq18}
\ee
where $F(M)$ is defined by (\ref{Eq6}), $L$ is a half-width
of a fault plane, and $\Gamma (. \, , \, .)$ is an incomplete
gamma function (Abramowitz and Stegun, 1972, Eq.~6.5.3).

The other possibility is to assume that earthquake centroid
density increases linearly with increasing depth up to the
middle of the fault width.
The density decreases to zero thereafter.
Kagan (2007, Fig.~6) shows that such a feature is a common
occurrence.
Then the survivor function would be
\be
\Psi_2 (M) \ \propto \ \int\limits_0^L F(M) \, \zeta \, d
\zeta \ = \ \left ( M_t \over M \right
)^{\beta} \, \int\limits_0^L \, \exp \left ( {{M_t \, - \, M}
\over {C \, \zeta^3} } \right ) \zeta \, d \zeta \ = \
\left ( M_t \over M \right )^{\beta} \,
{ { \Gamma \left ( - { 2 \over 3 }, \,
{ {M - M_t} \over {C \, L^3 } }
\right ) }
\over { 3 \left ( { C \over {M-M_t} } \right ) ^{2/3} } }
\, .
\label{Eq19}
\ee

\begin{figure}
\begin{center}
\includegraphics[width=0.75\textwidth]{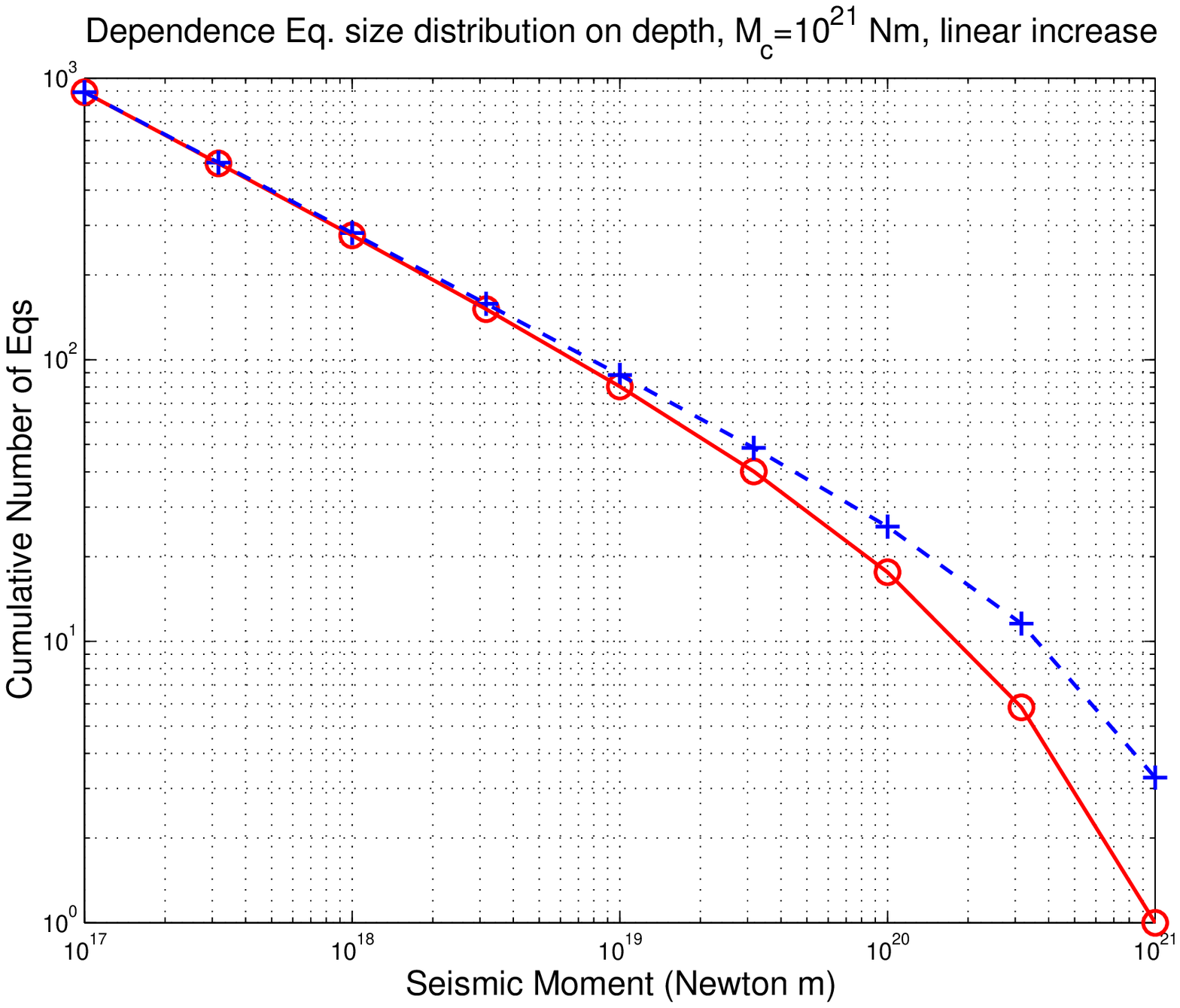}
\caption{\label{fig12}
}
\end{center}
Two theoretical moment-frequency curves: the
number of earthquakes with moment ($M$) larger than or equal
to $M$ as a function of $M$, moment threshold
$M_t=10^{17.0}$~Nm ($m_t=5.33$).
Dashed line shows tapered Gutenberg-Richter distribution: the
G-R law restricted at large seismic moments by an exponential
taper with the corner moment $10^{21.0}$ Nm ($m_c=8.0$).
The slope of the linear part of the curve corresponds to
$\beta=0.50$.
The solid line is a plot for Eq.~\ref{Eq18}, with a
half-width of a fault, $L=10$~km, and $C=1.0$.
The curves are normalized so that the solid line has an
ordinate 1.0 at the right-hand end.
\end{figure}

Fig.~\ref{fig12} displays two survivor functions corresponding
to Eqs.~\ref{Eq6} and \ref{Eq19}.
In a loglog plot the former function has a linear part for the
moment $M$ values that are significantly smaller than the
corner moment $M_c$.
We take a slope $\beta$ to be $0.5$.
The curve has an exponential taper for $M$ close to $M_c$.
On the other hand, the latter function (\ref{Eq19}) is slightly
convex even for small moment values.
It is formed by a sum of distributions similar to (\ref{Eq6})
but with the corner moment increasing from zero to the
maximum, $M_c$.
Therefore, we can only calculate an effective slope
($\beta^{\prime}$) of the curve; for the moment range
$10^{17}$---$10^{19}$~Nm, the slope $\beta^{\prime}$ in the
plot is 0.523.

We calculate three theoretical curves for both equations
(\ref{Eq18} and \ref{Eq19}):
(a) fault width $L=200$~km and $M_c=10^{23}$~Nm ($m_c=9.33$);
(b) $L=200$~km and $M_c=10^{22}$~Nm ($m_c=8.67$); and
(c) $L=20$~km and $M_c=10^{21}$~Nm ($m_c=8.0$).
For the formula (\ref{Eq18}) the $\beta^{\prime}$-values are
1.3, 2.8, and 6.6\% higher than the original $\beta$-value,
for the second formula these exponent increases are reduced by
a factor of 1.4.

\section{Discussion }
\label{disc}

In the previous sections we analyzed the index of the
power-law distribution for earthquake size (the $\beta$-value)
to argue that its true value is 1/2, the value suggested by
theoretical arguments.
The direct $\beta$ measurements for scalar seismic moment
based on catalog analyses (Kagan, 1999; Bird and Kagan, 2004)
usually yield a value in the range 0.63--0.67, equivalent to
the commonly known G-R $b$-value of 0.95--1.0.
Four systematic and random factors that bias the
$\beta$-value estimate upwards are investigated: dependence of
errors on the magnitude, earthquake sequences, complexity
of earthquake source, and a finite size of earthquake faults
(items 4-7 in the Introduction).
We found that these factors would decrease the observational
$\beta$-estimate by about 1--3\%, 5--10\%, 3--7\%, and 2--6\%,
respectively.
Of these values the third is most uncertain, because it is
based on extrapolating immediate aftershock focal mechanisms
to the mainshock' rupture time.

If we combine the above biases and apply them to the most
accurately determined $\beta$-value, i.e., $\beta$ equals from
0.63 to 0.64 for subduction zones (Bird and Kagan, 2004), the
corrected $\beta$-values would be on the order of 0.52--0.56.
It is quite feasible that the second and third correction term
are underestimated.
This would imply that $\beta$ is close to 0.5 and possibly
equals ${1 \over 2}$ exactly.


What theoretical conclusions could be drawn from this result?
Solid state physicists explain new results on the
scale-invariant distribution of dislocation avalanche size by
suggesting a new theoretical approach to crystal plasticity
(Zaiser, 2006).
According to this interpretation, at a micro-scale the crystal
deformation proceeds through intermittent bursts similar to
earthquakes.
Only at a larger, meso-scale does plastic deformation proceed
as a smooth, homogeneous, quasi-laminar flow process.
Crystal boundaries seem to influence this transition.
In a single crystal the power-law distribution for energy of
dislocation avalanches is observed at the scale range of
$10^6$.
In polycrystal materials, the power-law distribution of bursts
is also observed, but its size is limited by an upper cutoff.

Zaiser and Moretti (2005) and Csikor {\it et al.}\ (2007)
propose the following probability density function for the
dislocation avalanche energy or strain
\be
P(s) \ = \ C \, s^{-\tau} \,
\exp \left [ -(s/s_0)^2 \right ] \, ,
\label{Eq20}
\ee
where $C$ is a normalization constant, $\tau$ is a scaling
exponent ($\tau \simeq 1.5$), and $s_0$ is the characteristic
strain of the largest avalanches.
This formula is similar to our (\ref{Eq5}) with exponent
$\tau = 1 + \beta$, but the decay taper at large strain values
is Gaussian like.
We use instead the exponential decay.
Because statistics on the largest events are insufficient in
both cases, we cannot distinguish by observation between these
formulas.

Therefore, plastic, ductile deformation proceeds by two
very different mechanisms: (a) intermittent displacement at
micro-scale with scale-invariant distribution of strain steps
and the universal value of the power-law exponent
($\tau=1.5$), and (b) in contrast a smooth flow at larger
scales.
Because detailed quantitative observation at small sub-grain
scales was not possible until recently, the first mechanism
had been largely ignored (Zaiser, 2006, p.~241).

The above considerations can be supported to some degree by
recent analysis of earthquake size distribution.
Bird and Kagan (2004) found that the exponent $\beta$ of the
power-law distribution appears universal in all eight
tectonic provinces of global seismicity.
In stark contrast, the corner moment differs by many orders of
magnitude, from $10^{18}$ ($m_c=6.0$) for oceanic normal
faults to $10^{23.3}$~Nm ($m_c=9.5$) for subduction zones.

Kagan (2002a, pp.~538-9) proposed that the observed $b$-value
differences in volcanic areas, at creeping faults, and at the
boundary between brittle crust and plastic deformation in
the upper mantle may also be due to significantly varied
corner moments.
If earthquake populations with different $m_c$ are mixed, the
resulting statistical distribution could be interpreted as
belonging to a power-law with the exponent $\beta$
significantly exceeding 0.6--0.7, the value normally observed
in tectonic earthquakes.

Another geophysical phenomenon, the non-volcanic
seismic tremor (Schwartz and Rokosky, 2007; Beroza and Ide,
2009, and references therein), may be explained by the
same physical mechanism.
The tremor represents long duration (minutes to hours) of a
high-amplitude seismic signal, appearing similar to many small
concatenated earthquake signals (Shelly {\it et al.}, 2007).
The first observation of such non-volcanic tremors came
during the aftershock sequence of the April 26, 1966 Tashkent
(Uzbekistan) earthquake (Antsyferov {\it et al.}, 1971a,
1971b).

The tremor signals are sometimes quite pulsed in nature.
For example, the temporal cumulative plot of seismic moment
increase for tremors (Fig.~4d in Hiramatsu {\it et al.}, 2008)
look similar to curves of crystal micro-deformation due to
dislocation avalanches (Fig.~14 in Zaiser, 2006): both
diagrams show that the displacement increases in discrete
steps, each step is followed by a plateau.

Such tremors have been registered in diverse tectonic
environments recently (Japan, Cascadia, New Zealand, Costa
Rica, Taiwan, California).
Tremor and other slow-slip events are typically found on
the deep extension of faults, just below the region
that produces the more familiar, `ordinary' earthquakes.
This recent observation of tremors has resulted in a flurry of
research across many geologic and geophysical disciplines.

If tremors are a feature transitional between real earthquakes
and seismic signal bursts, they are similar to the dislocation
avalanches described above.
Both phenomena occur in conversion from the brittle to plastic
mode of solid deformation.
Seismic tremors, which are interpreted as small, continually
occurring earthquakes, may also have the same scale-invariant,
power-law features as earthquakes in brittle crust.
Hiramatsu {\it et al.}\ (2008) measured moment-frequency
relation for tremors and found that it can be approximated by
an exponential distribution rather than a power-law.
However, because of low signal-to-noise ratio for tremors,
only the upper tail of tremor size distribution can be
observed, and the upper tail for earthquake size distribution
(see Eq.~\ref{Eq5}) is also exponential (Kagan, 2002a; Bird and
Kagan, 2004).
The higher end of the size distribution for dislocation
avalanches also exhibits a non-power-law dependence
(\ref{Eq20}) which is close to exponential law.
We may conjecture that as with dislocation avalanches, the
size distribution of smaller tremor events would be the
power-law with the universal value of the exponent constant
($\beta=0.5$).
Further study of tremors should answer this question.

If we are correct about universality of the $\beta$-value
constant ($\beta=0.5$), the observed variations in the
$b$-parameter result from systematic and random effects not
properly accounted for (see the Introduction, items 1-3).
Therefore, all attempts to connect $\beta$-value
variability with various physical parameters and conditions
are eventually bound to fail.
However, studying the $b$- or $\beta$-values in local and
regional earthquake catalogs may still be useful, especially
if such investigations are needed to evaluate seismic hazard
and seismicity forecasts that would be prospectively tested
with the same catalogs.
In addition, when seismic activity or earthquake productivity
level is calculated for large earthquakes, the regular
$\beta$-estimates can be used.

If the hypothesis about the $\beta$-value constancy is
correct, we should investigate spatial changes in the corner
moment that seem to explain major modifications in the
deformation processes in solids.
Bird {\it et al.}\ (2009) showed that the relation
between relative plate velocity and seismicity is non-linear
for several types of plate boundaries.
Can the change in corner moment explain some of these
non-linearities?

If the hypothesis that the power-law exponent is a universal
constant and the corner moment is variable is correct, then it
would provide a new theoretical approach to features of
earthquake occurrence and account for the transition from
brittle to plastic deformation.
More extensive investigation of corner moment behavior may
afford new insight into regular earthquake occurrence and
recently discovered slow deformation and seismic tremor events
at the brittle-plastic crust boundary.
As often happens in complex systems, new laws and features may
be found to illuminate the transition from brittle fracture to
plastic flow.

\subsection* {Acknowledgments
}
\label{Ackn}
I am grateful to Dave Jackson, Paul Davis, Peter Bird,
Karin Dahmen, and especially to Jeremy Zechar for useful
discussion and suggestions.
I thank Kathleen Jackson for significant improvements in the
text.
The author appreciates support from the National Science
Foundation through grant EAR-0711515, as well as from the
Southern California Earthquake Center (SCEC).
SCEC is funded by NSF Cooperative Agreement EAR-0529922 and
USGS Cooperative Agreement 07HQAG0008.
Publication 0000, SCEC.

\pagebreak

\centerline { {\sc References} }
\vskip 0.1in
\parskip 1pt
\parindent=1mm
\def\reference{\hangindent=22pt\hangafter=1}

\reference
Abramowitz, M. and Stegun, I.~A., (Eds.), 1972.
{\sl Handbook of Mathematical Functions}, Dover, NY, pp.~1046.

\reference
Antsyferov, M. S., Y. Y. Kagan, and N. G. Antsyferova,
1971a. Seismoacoustical studies of aftershocks, in: {\sl
Tashkent Earthquake, April 26, 1966}, Fan, Tashkent,
pp.~154-163 (in Russian);
({\cyr TASHKENT\cydot SKOE ZEMLETRYASENIE 26 aprelya 1966
goda}), see
\hfil\break
http://moho.ess.ucla.edu/$\sim$kagan/Tashkent\_1966.pdf

\reference
Antsyferov, M. S., N. G. Antsyferova, and Y. Y. Kagan,
1971b. {\sl Seismoacoustical Studies and the Problem of
Prediction of Dynamic Events}, Nauka, Moscow, pp.\ 136 (in
Russian);
({\cyr Antsyferov, M. S., N. G. Antsyferova, i Ya.\ Ya.\
Kagan, 1971. Se\i2smo\-aku\-sti\-che\-skie Issledovaniya i
Problema Prognoza Dinamicheskih Yavleni\i2}),
pp.~115-123 contain a report on these measurements, see
http://moho.ess.ucla.edu/$\sim$kagan/Tashkent\_AAK.pdf

\reference
Beroza, G. C., and S. Ide, 2009.
Deep Tremors and Slow Quakes,
{\sl Science}, {\bf 324}(5930), 1025-1026,
DOI: 10.1126/science.1171231.

\reference
Bird, P., and Y. Y. Kagan, 2004.
Plate-tectonic analysis of shallow seismicity: apparent
boundary width, beta, corner magnitude, coupled
lithosphere thickness, and coupling in seven tectonic settings,
{\sl Bull.\ Seismol.\ Soc.\ Amer.}, {\bf 94}(6), 2380-2399
(plus electronic supplement).

\reference
Bird, P., Y. Y. Kagan, D. D. Jackson, F. P. Schoenberg,
and M. J. Werner, 2009.
Linear and Nonlinear Relations Between Relative Plate Velocity
and Seismicity,
{\sl Bull.\ Seismol.\ Soc.\ Amer.}, {\bf 99}(6), 3097-3113
(plus electronic supplement).

\reference
Boettcher, M., and T. H. Jordan, 2004.
Earthquake scaling relations for mid-ocean ridge transform
faults,
{\sl J. Geophys.\ Res.}, {\bf 109}(B12), Art.\ No.\ B12302,
doi:10.1029/2004JB003110.

\reference
Boettcher, M. S., A. McGarr, and M. Johnston, 2009.
Extension of Gutenberg-Richter distribution to $M_W \ -1.3$, no
lower limit in sight,
{\sl Geophys. Res. Lett.}, {\bf 36}, L10307, doi:
10.1029/2009GL038080.

\reference
Clauset, A., Shalizi, C. R., and Newman, M. E. J., 2009.
Power-law distributions in empirical data,
{\sl SIAM Rev.}, {\bf 51}, 661-703.

\reference
Csikor, F. F., C. Motz, D. Weygand, M. Zaiser, S. Zapperi,
2007.
Dislocation Avalanches, Strain Bursts, and the Problem of
Plastic Forming at the Micrometer Scale,
{\sl Science}, {\bf 318}(5848), 251-254.

\reference
Dahmen, K. A., Y. Ben-Zion, and J. T. Uhl, 2009.
Micromechanical Model for Deformation in Solids with Universal
Predictions for Stress-Strain Curves and Slip Avalanches,
{\sl Phys.\ Rev.\ Lett.}, {\bf 102}(17), Article Number:
175501.

\reference
Dimiduk, D. M., Woodward, C., LeSar, R., Uchic, M. D., 2006.
Scale-Free Intermittent Flow in Crystal Plasticity,
{\sl Science}, {\bf 312}, 1188-1190.

\reference
Dziewonski, A.\ M., Chou, T.-A., and Woodhouse, J.\ H., 1981.
Determination of earthquake source parameters from waveform
data for studies of global and regional seismicity,
{\sl J.\ Geophys.\ Res.}, {\bf 86}, 2825-2852.

\reference
Dziewonski, A.\ M., and Woodhouse, J.\ H., 1983.
An experiment in systematic study of global seismicity:
centroid-moment tensor solutions for 201 moderate and large
earthquakes of 1981,
{\sl J.\ Geophys.\ Res.}, {\bf 88}, 3247-3271.

\reference
Ekstr\"om, G., 2007.
Global Seismicity: Results from Systematic Waveform Analyses,
1976-2005,
in {\sl Treatise on Geophysics}, {\bf 4}(4.16), ed.\ H.
Kanamori, pp.~473-481.

\reference
Ekstr\"om, G., A. M. Dziewonski, N. N. Maternovskaya
and M. Nettles, 2005.
Global seismicity of 2003: Centroid-moment-tensor solutions
for 1087 earthquakes,
{\sl Phys.\ Earth planet.\ Inter.}, {\bf 148}(2-4), 327-351.

\reference
Enescu, B., J. Mori, M. Miyazawa, and Y. Kano, 2009.
Omori-Utsu Law $c$-Values Associated with Recent
Moderate Earthquakes in Japan,
{\sl Bull.\ Seismol.\ Soc.\ Amer.}, {\bf 99}(2A), 884-891.

\reference
Feller, W., 1966.
{\sl An Introduction to Probability Theory and its
Applications}, {\bf 2}, J. Wiley, New York, 626~pp.

\reference
Frohlich, C. and Davis, S.D., 1999.
How well constrained are well-constrained T, B, and P axes in
moment tensor catalogs?,
{\sl J.\ geophys.\ Res.}, {\bf 104}, 4901-4910.

\reference
Frohlich, C. and L. R. Wetzel, 2007.
Comparison of seismic moment release rates along different
types of plate boundaries,
Geophys. J. Int., {\bf 171}(2), 909-920,
doi: 10.1111/j.1365-246X.2007.03550.x

\reference
Geist, E.L., and T. Parsons, 2005.
Triggering of tsunamigenic aftershocks from large strike-slip
earthquakes: Analysis of the November 2000 New Ireland
earthquake sequence,
{\sl Geochemistry Geophysics Geosystems}, {\bf 6},
Article Number: Q10005.

\reference
Gutenberg, B. and Richter, C.F., 1944.
Frequency of earthquakes in California,
{\sl Bull.\ seism.\ Soc.\ Am.}, {\bf 34}, 185-188.

\reference
Harris, T.\ E., 1963.
{\sl The Theory of Branching Processes}, Springer, New York,
230~pp.

\reference
Hiramatsu, Y., T. Watanabe, and K. Obara, 2008.
Deep low-frequency tremors as a proxy for slip monitoring at
plate interface,
{\sl Geophys.\ Res.\ Lett.}, {\bf 35}, L13304,
doi:10.1029/2008GL034342.

\reference
Kagan, Y.~Y., 1982.
Stochastic model of earthquake fault geometry,
{\sl Geophys.\ J.\ Roy.\ astr.\ Soc.}, {\bf 71}(3), 659-691.

\reference
Kagan, Y.~Y., 1991a.
Seismic moment distribution,
{\sl Geophys.\ J. Int.}, {\bf 106}(1), 123-134.

\reference
Kagan, Y.~Y., 1991b.
Likelihood analysis of earthquake catalogues,
{\sl Geophys.\ J. Int.}, {\bf 106}(1), 135-148.

\reference
Kagan, Y.~Y., 1992.
On the geometry of an earthquake fault system,
{\sl Phys.\ Earth Planet.\ Inter.}, {\bf 71}(1-2),
15-35.

\reference
Kagan, Y.~Y., 1999.
Universality of the seismic moment-frequency relation,
{\sl Pure Appl.\ Geoph.}, {\bf 155}(2-4), 537-573.

\reference
Kagan, Y. Y., 2000.
Temporal correlations of earthquake focal mechanisms,
{\sl Geophys.\ J. Int.}, {\bf 143}(3), 881-897.

\reference
Kagan, Y. Y., 2002a.
Seismic moment distribution revisited: I. Statistical results,
{\sl Geophys.\ J. Int.}, {\bf 148}(3), 520-541.

\reference
Kagan, Y. Y., 2002b.
Aftershock zone scaling,
{\sl Bull.\ Seismol.\ Soc.\ Amer.}, {\bf 92}(2), 641-655.

\reference
Kagan, Y. Y., 2003.
Accuracy of modern global earthquake catalogs,
{\sl Phys.\ Earth Planet.\ Inter.\ (PEPI)}, {\bf 135}(2-3),
173-209.

\reference
Kagan, Y. Y., 2004.
Short-term properties of earthquake catalogs and models of
earthquake source,
{\sl Bull.\ Seismol.\ Soc.\ Amer.}, {\bf 94}(4), 1207-1228.

\reference
Kagan, Y. Y., 2007.
Earthquake spatial distribution: the correlation dimension,
{\sl Geophys.\ J. Int.}, {\bf 168}(3), 1175-1194.

\reference
Kagan, Y. Y., 2009.
On the geometric complexity of earthquake focal
zone and fault systems: A statistical study,
{\sl Phys.\ Earth Planet.\ Inter.}, {\bf 173}(3-4), 254-268,
DOI: 10.1016/j.pepi.2009.01.006.

\reference
Kagan, Y. Y., 2010.
Statistical distributions of earthquake numbers:
consequence of branching process,
{\sl Geophys.\ J. Int.}, {\bf 180}(3), 1313-1328.
doi: 10.1111/j.1365-246X.2009.04487.x

\reference
Kagan, Y. Y., P. Bird, and D. D. Jackson, 2010.
Earthquake Patterns in Diverse Tectonic Zones of
the Globe,
{\sl Pure Appl.\ Geoph.} ({\sl The Frank Evison Volume}), {\bf
167}(6/7), in press,
DOI: 10.1007/s00024-010-0075-3, \
\hfil\break
http://scec.ess.ucla.edu/$\sim$ykagan/globe\_index.html .

\reference
Kagan, Y.~Y., and L. Knopoff, 1981.
Stochastic synthesis of earthquake catalogs,
{\sl J.\ Geophys.\ Res.}, {\bf 86}(B4), 2853-2862.

\reference
Kanamori, H., 1977.
The energy release in great earthquakes,
{\sl J. Geophys.\ Res.}, {\bf 82}, 2981-2987.

\reference
Miguel, M.-C., A. Vespignani, S. Zapperi, J. Weiss and J.-R.
Grasso, 2001.
Intermittent dislocation flow in viscoplastic deformation,
{\sl Nature}, {\bf 410}, 667-671, doi:10.1038/35070524.

\reference
Molchan, G.\ M., and V.\ M.\ Podgaetskaya, 1973.
Parameters of global seismicity, I,
in: Keilis-Borok, V.\ I., (ed.), {\sl Computational
Seismology}, {\bf 6}, Nauka, Moscow, 44-66, (in Russian).

\reference
Okal, E.\ A., and B.\ A.\ Romanowicz, 1994.
On the variation of $b$-values with earthquake size,
{\sl Phys.\ Earth Planet.\ Inter.}, {\bf 87}, 55-76.

\reference
Otter, R., 1949.
The Multiplicative Process,
{\sl Annals Math.\ Statistics}, {\bf 20}(2), 206-224.

\reference
Park, S.C., and J. Mori, 2007.
Triggering of earthquakes during the 2000 Papua New Guinea
earthquake sequence,
{\sl J. Geophys.\ Res.}, {\bf 112}(B3), Article Number:
B03302.

\reference
Perez-Campos, X., S.K. Singh, and G.C. Beroza, 2003.
Reconciling teleseismic and regional estimates of seismic
energy,
{\sl Bull.\ Seismol.\ Soc.\ Amer.}, {\bf 93}(5), 2123-2130.

\reference
Rhoades, D.\ A., 1996.
Estimation of the Gutenberg-Richter relation allowing for
individual earthquake magnitude uncertainties,
{\sl Tectonophysics}, {\bf 258}, 71-83.

\reference
Rhoades, D.\ A., 1997.
Estimation of attenuation relations for strong-motion data
allowing for individual earthquake magnitude uncertainties,
{\sl Bull.\ Seismol.\ Soc.\ Amer.}, {\bf 87}(6), 1674-1678.

\reference
Rhoades, D.A., and Dowrick, D.J., 2000.
Effects of magnitude uncertainties on seismic hazard
estimates,
Proceedings of the 12th World Conference on Earthquake
Engineering, Auckland, New Zealand, 30th January - 4th
February 2000, Paper No.\ 1179. New Zealand Society for
Earthquake Engineering, Upper Hutt, New Zealand,
{\sl Bull.\ N.Z. Soc.\ Earthqu.\ Eng.}, {\bf 33}(3).

\reference
Richeton, T,. J. Weiss and F. Louchet, 2005.
Breakdown of avalanche critical behaviour
in polycrystalline plasticity,
{\sl Nature Materials}, {\bf 4}, 465-469.

\reference
Schorlemmer D., S. Wiemer, and M. Wyss, 2005.
Variations in earthquake-size distribution across different
stress regimes,
{\sl Nature}, {\bf 437}, 539-542.

\reference
Schwartz, S.Y., and J.M. Rokosky, 2007.
Slow slip events and seismic tremor at circum-pacific
subduction zones,
{\sl Reviews Geophysics}, {\bf 45}(3), Art. No. RG3004

\reference
Shelly, D. R., G. C. Beroza, and S. Ide, 2007.
Non-volcanic tremor and low frequency earthquake swarms,
{\sl Nature}, {\bf 446}, 305-307.

\reference
Thomson Scientific/ISI http://www.isinet.com/
ISI Web of Science, 2006.
The Thomson Corporation,
Thomson Reuters' Web of Science
\hfil\break
http://portal.isiknowledge.com/portal.cgi
(last accessed July 2009).

\reference
Tinti, S., and F. Mulargia, 1985.
Effects of magnitude uncertainties on estimating the
parameters in the Gutenberg-Richter frequency-magnitude law,
{\sl Bull.\ Seismol.\ Soc.\ Amer.}, {\bf 75}, 1681-1697.

\reference
Vere-Jones, D., 1976.
A branching model for crack propagation,
{\sl Pure Appl.\ Geophys.\ } ({\scaps pageoph}),
{\bf 114}, 711-725.

\reference
Vere-Jones, D., 1977.
Statistical theories of crack propagation,
{\sl Math.\ Geol.}, {\bf 9}, 455-481.

\reference
Vere-Jones, D., 1988.
Statistical aspects of the analysis of historical earthquake
catalogues,
in: C. Margottini, ed., {\sl Historical Seismicity of
Central-Eastern Mediterranean Region}, pp.~271-295.

\reference
Weiss, J., and D. Marsan, 2003.
Three-dimensional mapping of dislocation avalanches:
clustering and space/time coupling,
{\sl Science}, {\bf 299}, 89-92.

\reference
Wolfram, S., 1999.
{\sl The Mathematica Book},
4th ed., Champaign, IL, Wolfram Media, Cambridge, New York,
Cambridge University Press, pp.~1470.

\reference
Zaiser, M., 2006.
Scale invariance in plastic flow of crystalline solids,
{\sl Advances Physics}, {\bf 55}(1-2), 185-245.

\reference
Zaiser, M., and P. Moretti, 2005.
Fluctuation phenomena in crystal
plasticity -- a continuum model,
{\sl J. Statistical Mechanics},
doi:10.1088/1742-5468/2005/08/P08004.

\reference
Zaiser, M., Schwerdtfeger, J., Schneider, A. S., Frick, C. P.,
Clark, B. G., Gruber, P. A. and Arzt, E., 2008.
Strain bursts in plastically deforming molybdenum micro- and
nanopillars,
{\sl Philosophical Magazine}, {\bf 88}(30), 3861-3874.

\reference
Zhuang, J., Y. Ogata, and D. Vere-Jones, 2002.
Stochastic declustering of space-time earthquake occurrences,
{\sl J.\ Amer.\ Statist.\ Assoc.\ (JASA)}, {\bf 97}, 369-380.

\reference
Zhuang, J. C., Y. Ogata, and D. Vere-Jones, 2004.
Analyzing earthquake clustering features by using stochastic
reconstruction,
{\sl J. Geophys.\ Res.}, {\bf 109}(B5), Art.\ No.\ B05301.

\newpage

\begin{table}
\caption{Parameter values for relative errors.
}
\vspace{10pt}
\label{Table1}
\begin{tabular}{@{}lrrrrrr}
\hline
& & & & & & \\[-15pt]
Approximation
& $n$
& $c_0$
& $c_1$
& $c_2$
& $\rho$
& $\sigma$
\\
& & & & & & \\
\hline
& & & & & & \\
& \multispan4 CMT 1977-2008, $m_t=5.6$ \\
Linear & 8508 & $-1.20$ & $-0.352$ & & $-0.47$ & $0.274$ \\
Quadratic & 8508 & $-1.23$ & $-0.513$ & $0.198$ & $-0.47$ & $0.267$ \\
& & & & & \\
& \multispan4 CMT 1982-2008, $m_t=5.4$ \\
Linear & 11600 & $-1.21$ & $-0.447$ & & $-0.57$ & $0.274$ \\
Quadratic & 11600 & $-1.26$ & $-0.537$ & $0.208$ & $-0.57$ & $0.266$ \\
\\
\hline
\end{tabular}

\bigskip
$n$, the number
of $ m \ge m_t$ events;
for $c_0$, $c_1$, and $c_2$ see (\ref{Eq10});
\hfil\break
$\rho$, correlation coefficient;
$\sigma$, standard deviation of fit.
\hfil\break
For the linear approximation, the $c_2$ parameter is shown as
a dash.
\end{table}

\clearpage

\newpage

\begin{table}
\caption{
The $\beta$-values for various subdivisions of CMT catalog.
}
\vspace{10pt}
\label{Table2}
\begin{tabular}{lccccc}
\hline
& & & & & \\[-15pt]
\multicolumn{1}{c}{\#}&
\multicolumn{1}{c}{Earthquakes}&
\multicolumn{2}{c}{$m_t = 5.6$ }&
\multicolumn{2}{c}{$m_t = 5.8$ }
\\[2pt]
\multicolumn{1}{c}{}&
\multicolumn{1}{c}{}&
\multicolumn{1}{c}{$\beta$ }&
\multicolumn{1}{c}{Eq. \# }&
\multicolumn{1}{c}{$\beta$ }&
\multicolumn{1}{c}{Eq. \# }
\\[2pt]
\multicolumn{1}{c}{1}&
\multicolumn{1}{c}{2}&
\multicolumn{1}{c}{3}&
\multicolumn{1}{c}{4}&
\multicolumn{1}{c}{5}&
\multicolumn{1}{c}{6}
\\[2pt]
\hline
& & & & & \\[-15pt]
1. & Global & 0.6773 & 7369 & 0.6820 & 5450 \\
2. & Global declustered & 0.6480 & 5841 & 0.6568 & 4498 \\
3. & Global aftershocks included & 0.6229 & 5605 & 0.6366 & 4358 \\
4. & Trenches (Subduction zones) & 0.6463 & 4805 & 0.6507 & 3223 \\
& & & & & \\[-15pt]
\hline
\end{tabular}

\bigskip

Shallow earthquakes in
1977-2008 ($m_t = 5.8$)
and 1982-2008 ($m_t = 5.6$)
CMT catalog.

\hfil\break
\vspace{5pt}
\end{table}

\newpage


\parindent=0mm

\end{document}